\documentclass[a4paper,12pt]{article}
\usepackage[margin=1.3cm]{geometry}
\usepackage{comment}
\usepackage{amssymb,extarrows,graphicx,subfigure,setspace}
\usepackage{cite}
\usepackage{slashed}
\usepackage{tensor}
\usepackage[toc,page]{appendix}
\usepackage{color}
\usepackage{physics}
\usepackage{hyperref}
\usepackage{dirtytalk}
\hypersetup{colorlinks=true, linkcolor=blue, citecolor=red, linktoc=page}
\makeatother

\usepackage{float}
\usepackage{textcomp}
\usepackage{amsmath}
\newmuskip\pFqmuskip

\newcommand*\pFq[6][8]{%
  \begingroup 
  \pFqmuskip=#1mu\relax
  \mathcode`\,=\string"8000
  \begingroup\lccode`\~=`\,
  \lowercase{\endgroup\let~}\pFqcomma
  {}_{#2}F_{#3}{\left[\genfrac..{0pt}{}{#4}{#5};#6\right]}%
  \endgroup
}
\newcommand{\pFqcomma}{\mskip\pFqmuskip}

\usepackage{mathrsfs}
\usepackage{hyperref}

\newcommand{\be}{\begin{equation}}
\newcommand{\bea}{\begin{eqnarray}}
\newcommand{\eea}{\end{eqnarray}}
\newcommand{\ba}{\begin{array}}
\newcommand{\ea}{\end{array}}
\newcommand{\ee}{\end{equation}}
\newcommand{\bes}{\begin{equation*}}
\newcommand{\beas}{\begin{eqnarray*}}
\newcommand{\eeas}{\end{eqnarray*}}
\newcommand{\bas}{\begin{array*}}
\newcommand{\eas}{\end{array*}}
\newcommand{\ees}{\end{equation*}}

\numberwithin{equation}{section}

\begin{document}

\begin{center}
\Large{\bf Mutual Influence of Photon Sphere and Non-Commutative Parameter in Various Non-Commutative Black Holes: Part I- Towards evidence for WGC}\\
 \small \vspace{0.5cm}
 {\bf  Mohammad Ali S. Afshar $^{\star}$\footnote {Email:~~~m.a.s.afshar@gmail.com}}, \quad
 {\bf Jafar Sadeghi $^{\star}$\footnote {Email:~~~pouriya@ipm.ir}}\\
\vspace{0.5cm}$^{\star}${Department of Physics, Faculty of Basic
Sciences,\\
University of Mazandaran
P. O. Box 47416-95447, Babolsar, Iran}\\
\small \vspace{0.5cm}
\end{center}
\begin{abstract}
Non-commutative black holes(NCBH), due to the non-commutativity of spacetime coordinates, lead to a modification of the spacetime metric. By replacing the Dirac delta function with a Gaussian distribution, the mass is effectively smeared, eliminating point-like singularities. Our objective is to investigate the impact of this change on spacetime geodesics, including photon spheres and time-like orbits. We will demonstrate how the photon sphere can serve as a tool to classify spacetime, illustrating the influence of the NC parameter and constraining its values in various modes of these black holes. Additionally, using this classification, we will show how the addition of the nonlinear Einstein-Born-Infeld(BI) field to the model enhances its physical alignment with reality compared to the charged model. In the dS BI model, we will show how the study of the effective potential and photon sphere can provide insights into the initial structural status of the model, thereby establishing this potential as an effective tool for examining the initial conditions of black holes. Finally, by examining super-extremality conditions, we will show that the AdS BI model, with the necessary conditions, can be a suitable candidate for studying and observing the effects of the Weak Gravity Conjecture (WGC).
\end{abstract}
\tableofcontents
\section{Introduction}
In the past, there were suspicions regarding the practical existence of ultra-compact gravitational structures, particularly black holes, which led to their classification as theoretical concepts. However, numerous recent observational data have confirmed the presence of structures with behavioral patterns very similar to ultra-compact gravitational entities in the vast expanse of the sky. These structures play a significant role in shaping and maintaining the current equilibrium of the cosmos.
Despite this, the current scientific and instrumental limitations, coupled with the vastness of the universe, have prevented us from formulating a unified or precise mathematical model for these structures. However, by relying on principles such as mathematical induction and recurring fractal patterns in creation, scientists have attempted to combine the behavior of known fields with observational and cosmic effects. The goal of this approach is to develop and expand models that, in the future, will allow us to select the most accurate and superior model that best aligns with creation. Consequently, various black holes have been introduced to date, combining gravity with different geometric and electromagnetic perspectives.\\
One example of such developed models is black holes with NC structures. The application of NC geometry to black holes began to gain traction in the early 21st century \cite{1,2,3}. Researchers sought to understand how quantum gravity effects could alter the properties of black holes. The concept of NCBH is rooted in the broader framework of NC geometry, a mathematical approach that modifies the classical notion of spacetime. This concept arises from the intersection of quantum mechanics and general relativity, aiming to address the singularities and infinities that plague classical black hole solutions. The idea of NC geometry, where spacetime coordinates do not commute, was introduced as a way to incorporate quantum effects into the fabric of spacetime itself. NCBH differ from their classical counterparts in several key ways. The most significant difference lies in the modification of the spacetime metric due to non-commutativity. In classical general relativity, the spacetime around a black hole is described by the Schwarzschild metric (for non-rotating black holes) or the Kerr metric (for rotating black holes). However, in a NC framework, these metrics are altered to incorporate the effects of a minimal length scale. The resulting metrics replace the Dirac delta function with a Gaussian distribution, effectively spreading out the mass and eliminating point-like singularities. The smearing of the mass distribution affects the emission spectrum, leading to potential observational signatures that could distinguish NCBH from their classical counterparts. Also, unlike classical black holes that evaporate completely through Hawking radiation, NCBH predict the existence of stable remnants. These remnants are characterized by a zero-temperature final state, preventing the complete evaporation and providing a potential solution to the information loss paradox \cite{4,5,6}.\\
While NCBH models offer several advantages, which led to be studied in different forms \cite{7,8,9,10}, they also come with certain disadvantages compared to other models.  For instance, NC geometry introduces additional mathematical complexity that making the models harder to work with and understand, or some NCBH models predict negative radial and tangential pressures, which are difficult to explain physically \cite{11}.\\
After constructing an initial model based on the equations of general relativity, which primarily focuses on preserving the physical, energetic, and geometric aspects of the model, the next crucial step is to analyze the behavior of the constructed model in interaction with its surroundings. This means examining the gravitational, optical, or thermodynamic behavior patterns exhibited by the model.
In this article, we will employ a topological method based on the analysis of winding numbers and topological charge. This mathematical approach, which has recently garnered significant attention, extends beyond the study of photon spheres and photon rings to thermodynamics as well \cite{12,13,14,15,16,17,18,19,20,21,22,23,23.1,24,25,26,27,28,29}.
We know that, based on topological reasoning, in a stationary, axisymmetric, four-dimensional ultra-compact object without a horizon, light rings always come in pairs, one stable and the other unstable \cite{30}. On the other hand, we know that stationary, axisymmetric, asymptotically flat, four-dimensional black holes always have at least one unstable light ring (LR) \cite{31}. Additionally, studies have shown that at least one standard photon sphere exists outside the black hole not only in asymptotically flat spacetimes but also in AdS and dS spacetimes \cite{32}. Given the above and considering the gravitational equation patterns, it seems that the existence of a photon sphere is one of the requirements for ultra-compact gravitational structures. Accordingly, in our previous two works, we have shown that there is often a reciprocal relationship between the structural parameters of the model and the photon sphere. This relationship allows us to classify the space around the model based on known parameters or, conversely, to determine the parameter range for each model based on the studied spatial region \cite{33,34,35}.\\
In this paper, in addition to studying the mutual influence between the appearance and location of the photon sphere and the effective range of the NC parameter in various NCBH modes, we will also examine the impact of this parameter on time-like orbits. Furthermore, by investigating the possibility of the emergence of super-extremal black holes in charged models, we will seek a suitable candidate for the WGC.\\ Based on the above information, we are motivated to structure the paper as follows:
In section 2, we will review the fundamentals and principles in two parts. In the first part, we will discuss Duan’s topological mapping and the effective potential, explaining how to calculate the topological charges for the photon sphere and then, we will address timelike geodesics and introduce the relevant equations.\\
In sections 3 and 4, we will examine and analyze the various introduced models using the methods discussed. Then in section 5, we will review intriguing points in the search for evidence of the WGC and discuss the potential of the photon sphere as an effective measurement tool, even in this context. Finally, in section 6, we will present the conclusions drawn from our analysis.
 \section{Methodology}
 In this section we will review the fundamentals and principles in two parts as follows:
\subsection{Topological Photon Sphere } 
A photon sphere is a null ring that serves as the lower bound for any stable orbit and demonstrates the extreme bending of light rays in a very strong gravitational field. It manifests in two forms: unstable, where small perturbations cause photons to either escape or fall into the black hole, and stable, where perturbations do not allow photons to escape. The unstable type is useful for determining black hole shadows, while the stable type induces spacetime instability.\\
In the traditional approach to studying photon spheres, it is essential to derive the Lagrangian from the action and subsequently construct the Hamiltonian. Once the Hamiltonian is determined, the effective potential can be constructed, allowing for the study of the photon sphere. In this context, the effective potential is a function of the particle's energy and angular momentum.\\In this work, instead of using the conventional method and traditional effective potential, we employ an effective potential introduced by Cunha and colleagues for spherically symmetric ultra-compact objects \cite{30,31}, which was later utilized by Shao-Wen Wei to study the photon sphere of 4-dimensional black holes in AdS and dS spacetimes \cite{32}. A detailed explanation of the calculations can be found in references \cite{32,33,34,35}.
The significant advantage of using this potential lies in its independence from the energy and angular momentum of the incoming particle, indicating that this rewritten effective potential is solely a function of the spacetime geometry. Additionally, the use of the equatorial plane for mapping the vector field $\phi$, similar to the Poincaré plane, and the resulting dimensional reduction is another notable benefit. Ultimately, the ability to study a broader range of the space surrounding an ultra-compact object is a fundamental aspect that, for the first time, leads to the classification of different spacetime regions based on this method.\\
In the first step, we consider a general vector field as  $\phi$  which can be decomposed into two components, $\phi^r$ and $\phi^\theta$,
\begin{equation}\label{1}
\phi=(\phi^{r}, \phi^{\theta}),
\end{equation}
 also here, we can  rewrite the vector as $\phi=||\phi||e^{i\Theta}$, where $||\phi||=\sqrt{\phi^a\phi^a}$, or $\phi = \phi^r + i\phi^\theta$.
 Based on this, the normalized vector is defined as,
 \begin{equation}\label{2}
n^a=\frac{\phi^a}{||\phi||},
\end{equation}
 where $a=1,2$  and  $(\phi^1=\phi^r)$ , $(\phi^2=\phi^\theta)$.
 Based on Noether's theorem, we know that charge and current can be written in the following form:
 \begin{center}
 $\partial_\nu j^\nu=0$
 \end{center}
and
\begin{equation}\label{3}
Q=\int_{\Omega}j^0d^{2}x,
\end{equation}
where $j^0$ is the charge density. With some calculations, the current can be rewritten in the following form \cite{32,33,34,35}:
\begin{equation}\label{(4)}
j^{\mu}=J^{\mu} \! \left(X \right) \delta^{2} \! \left(\phi \right),
\end{equation}
where $\delta$ is the Dirac delta function. By using the above relations the topological charge become,
\begin{equation}\label{5}
Q=\int_{\Omega}J^{0} \! \left(X \right) \delta^{2} \! \left(\phi \right)d^{2}x,
\end{equation}
where $X=\frac{\phi}{x}$.\\In the next stage, with respect to the necessity of spherical symmetry as a prerequisite for studying this method, and given the most general form of the metric in 4-dimensional form we have:
\begin{equation}\label{(6)}
\mathit{ds}^{2}=-\mathit{dt}^{2} f \! \left(r \right)+\frac{\mathit{dr}^{2}}{g \! \left(r \right)}+\left(d\theta^{2}+d\varphi^{2}
\sin \! \left(\theta \right)^{2}\right) h \! \left(r \right)=\mathit{dr}^{2} g_{\mathit{rr}}-\mathit{dt}^{2} g_{\mathit{tt}}+d\theta^{2} g_{\theta \theta}+d\varphi^{2} g_{\varphi \varphi},
\end{equation}
for the new form of effective potential,we have \cite{32}:
\begin{equation}\label{(7)}
\begin{split}
H(r, \theta)=\sqrt{\frac{-g_{tt}}{g_{\varphi\varphi}}}=\frac{1}{\sin\theta}\bigg(\frac{f(r)}{h(r)}\bigg)^{1/2}.
\end{split}
\end{equation}
Based on this potential, we now define the vector field vector field $\phi=(\phi^r,\phi^\theta)$ as follows:
\begin{equation}\label{(8)}
\begin{split}
&\phi^r=\frac{\partial_rH}{\sqrt{g_{rr}}}=\sqrt{g(r)}\partial_{r}H,\\
&\phi^\theta=\frac{\partial_\theta H}{\sqrt{g_{\theta\theta}}}=\frac{\partial_\theta H}{\sqrt{h(r)}}.
\end{split}
\end{equation}
We know that photon spheres typically reside at the extremum points of the effective potential function. Additionally, it is evident that the vector field $ \phi $ is a function of the derivatives of $ H(r)$. Given the properties of the delta function, it can be inferred that the charges defined in equation (2.3) are non-zero exclusively at locations where $ \phi $ equals zero.
Therefore, not only do the zeros of the radial component of $ \phi $ indicate the location of the photon sphere, but a topological charge ,Q , can also be assigned to each photon sphere.\\
In his paper, Wei demonstrated that by examining winding numbers and calculating total topological charges (TTC)for a model, a TTC value of -1 indicates black hole behavior, whereas a TTC value of 0 corresponds to a naked singularity \cite{32}. In our previous papers, in addition to using the above analysis and topological charge analysis to classify the behavior of the studied model, we also examined the geometric behavior of the effective potential. Our observations in several cases \cite{28,29,32,33,34,35} indicate that when the asymptotically AdS or flat system under study shows a TTC  of -1, it generally possesses an event horizon and exhibits black hole behavior. In such a scenario, the potential function shows a dominant maximum outside the event horizon, indicating instability at the location of the photon sphere. Conversely, when the ultra-compact object under study shows a TTC of 0 or 1, the system generally lacks an event horizon and takes the form of a naked singularity. In this case, the potential function shows a dominant minimum alongside a maximum, indicating the presence of an extra stable photon sphere.\\
Of course, The situation is somewhat different for dS structures. In dS models, one of the horizons that appears is called the cosmological horizon or the de Sitter horizon. This horizon typically appears so far away that it practically prevents the formation of any potential beyond it. In this case, we limit our study to the region between the Cauchy horizon (the smaller horizon) and the cosmological horizon. Our studies show that dS models often appear in the form of black holes, and the form of naked singularity may not be observable using this method.
\subsection{Time-like circular orbits (TCOs) } 
One of the prominent characteristics of gravitational fields is the specific behavioral patterns and boundaries they impose on the surrounding spacetime geometry. This imposition of order becomes significantly more intense in the case of ultra-compact structures, to the extent that in regions very close to these structures, we lose our classical geometric-causal interpretation. Consequently, we are compelled to conceal everything that "exists" but cannot be interpreted, behind the event horizon by using the weak cosmic censorship conjecture (WCCC). However, beyond the event horizon, our field equations still appear to be reasonably applicable, allowing us to accurately identify and classify these geometric-kinematic boundaries. Based on this, we have:\\
1. Null Geodesics (Light-like): The trajectories of massless particles and photons.\\
2. Time-like Geodesics: The trajectories of massive objects.\\
These geodesics can, depending on the characteristics of the gravitational potential model, either be without an observable turning point or possess a turning point and form loops. In null geodesics, these are referred to as light rings (photon spheres), and in time-like geodesics, they are known as TCO (Time-like Circular Orbits). Both types of orbits can be stable or unstable. While we have previously discussed light rings, here we will focus on the behavior of TCOs. Our assumptions in this study include a static, axi-symmetric spacetime with $Z_2$ symmetry in a 1+3 dimensional framework which with this Given symmetries, discussing in the equatorial plane will not compromise the comprehensiveness of the analysis. In order to better study and categorize the behavior of TCOs based on the metric equation ,Eq. (\ref{(6)}), We consider the following quantities \cite{36}:
\begin{equation}\label{(9)}
A =g_{\varphi \varphi} E^{2}+g_{\mathit{tt}} L^{2},
\end{equation}
\begin{equation}\label{(10)}
B =-g_{\varphi \varphi} g_{\mathit{tt}},
\end{equation}

which E,L are the energy and the angular momentum. Now the Lagrangian can be recast as:
\begin{equation}\label{(11)}
2 \mathfrak{L} =-\frac{A}{B}=\zeta,
\end{equation}
where $\zeta = -1, 0 $ for time-like, null  geodesics, respectively. Using the above Lagrangian, the effective potential can be rewritten as follows:
\begin{equation}\label{(12)}
V_{\mathit{eff} \! \left(\zeta \right)}\! \left(r \right)=\zeta +\frac{A}{B}.
\end{equation}
To have a more accurate analysis of the behavior of TCOs, it is better to express the concept of angular velocity (as measured by an observer at infinity) and $\beta$ in terms of metric parameters.
\begin{equation}\label{(13)}
\Omega_{\pm}=\frac{g_{\mathit{tt}} L}{g_{\varphi \varphi} E},
\end{equation}
\begin{equation}\label{(14)}
\beta_{\pm}=-A \! \left(r^{\mathit{cir}},\Omega_{\pm},\Omega_{\pm}\right),
\end{equation}
in which $\pm$ is a sign of prograde/retrograde orbits \cite{36}. Now we have the necessary relations for the behavioral study of TCOs, so we will analyze them according to these equations. We know that to have a circular orbit, we need turning points, which must satisfy the following two conditions:
\begin{equation}\label{(15)}
\left(V_{-1}=0\right)_{\mathit{cir}},
\end{equation}
\begin{equation}\label{(16)}
\left(\mathrm{D}\! \left(V_{-1}\right)\! \left(r \right)=0\right)_{\mathit{cir}}.
\end{equation}
However, given the complexity of the models, these two conditions are certainly not sufficient to determine the status of TCOs. For instance, when the parameter $\beta$ in Eq. (\ref{(14)}) becomes negative, it results in negative energy and angular momentum, leading to an imaginary angular velocity in Eq. (\ref{(13)}) which does not have any physical interpretation. Therefore, the sign of parameter $\beta$ plays a crucial role in determining the permissible and impermissible regions for the presence of TCOs.\\
\begin{equation}\label{(17)}
\begin{split}
& \left(0<\mathrm{D}^{\left(2\right)}\! \left(V_{-1}\right)\! \left(r \right)\right)_{\mathit{cir}}\Rrightarrow Unstable(UTCO),\\&\left(\mathrm{D}^{\left(2\right)}\! \left(V_{-1}\right)\! \left(r \right)<0\right)_{\mathit{cir}}\Rrightarrow Stable(STCO).\\
\end{split}
\end{equation}
The transition from UTCOs to STCOs and vice versa, given the continuity of the functions under study, must occur across a boundary known as the Marginally Stable Circular Orbit (MSCO). An MSCO should be understood as the stable circular orbit with the smallest radius, which is continuously connected to spatial infinity by a set of stable TCOs \cite{36}.
\section{Charged Non-Commutative Black Hole}
In this section, we first consider a simpler model that includes only electric charge. Our initial and straightforward objective is to examine the topological photon sphere and the stable and unstable TCOs of the model. However, our true aim is to investigate several distinct aspects. Firstly, we intend to explore the mutual influence of the NC parameter on the location of the photon sphere and, conversely, the constraints imposed by the presence of the photon sphere on the effective range of this parameter. Additionally, we aim to determine whether the pattern proposed in \cite{36} for TCOs around the light ring holds true in this model. Furthermore, we seek to ascertain whether this model can represent a black hole in a super-extremal form, allowing us to search for evidence supporting the WGC. Finally, in the next section, when additional parameters such as the BI parameter are incorporated into the model, we aim to more clearly examine the impact of these added parameters on the constraints. 
Because comparing the results of the two models can provide better insights into how these parameters influence.\\
The metric for such black hole is \cite{37,38}:
\begin{equation}\label{(18)}
f =1-\frac{4 \gamma \! \left(\frac{3}{2},\frac{r^{2}}{4 \Xi}\right) m}{\sqrt{\pi}\, r}+\frac{\left(\gamma \! \left(\frac{1}{2},\frac{r^{2}}{4 \Xi}\right)^{2}-\frac{\gamma \left(\frac{1}{2},\frac{r^{2}}{2 \Xi}\right) r \sqrt{2}}{2 \sqrt{\Xi}}+\gamma \! \left(\frac{3}{2},\frac{r^{2}}{4 \Xi}\right) r \sqrt{2}\, \sqrt{\frac{1}{\Xi}}\right) q^{2}}{r^{2} \pi}
\end{equation}
where $\Xi$ is NCe parameter, m is the total mass of the source,q is the total electric charge and the $\gamma$ is the lower incomplete Gamma function which can be written in terms of $\Gamma$ (the upper incomplete Gamma function) in the following form:
\begin{equation}\label{(19)}
 \gamma (\frac{3}{2},\frac{r^{2}}{4\,\Xi})\equiv \int_{0}^{\frac{r^{2}}{4\,\Xi}}\sqrt{t}\times {\mathrm e}^{t}\quad d t =\Gamma (\frac{3}{2})-\Gamma (\frac{3}{2},\frac{r^{2}}{4\,\Xi}),
\end{equation}
As stated before, one of the main features of NCBH is the smearing of the mass distribution, meaning that the mass is spread over a finite region rather than being concentrated at a single point. This alteration, among other effects, also impacts the structure of the black hole's event horizon, with the parameter $\Xi$ playing a significant role in determining the event horizon. For instance, as illustrated in Fig(1), beyond a certain value of $\Xi$, the black hole structure is effectively lost, transforming into a naked singularity. Before continuing the discussion, it is important to note that due to the complexities of the model, an analytical solution to the metric function in its general form is very difficult. Therefore, we must proceed with our analysis in a numerical form. For instance, by selecting $ m = 1$ and $ q = 0.6 $, it can be observed that, firstly, the structure does not result in a black hole for any value of $ \Xi $. Secondly, although smaller values of $ \Xi $ yield more physically meaningful results, but the structure retains the event horizon up to a value of $\Xi = 0.2184$.
\begin{figure}[H]
 \begin{center}
 \subfigure[]{
 \includegraphics[height=6.5cm,width=8cm]{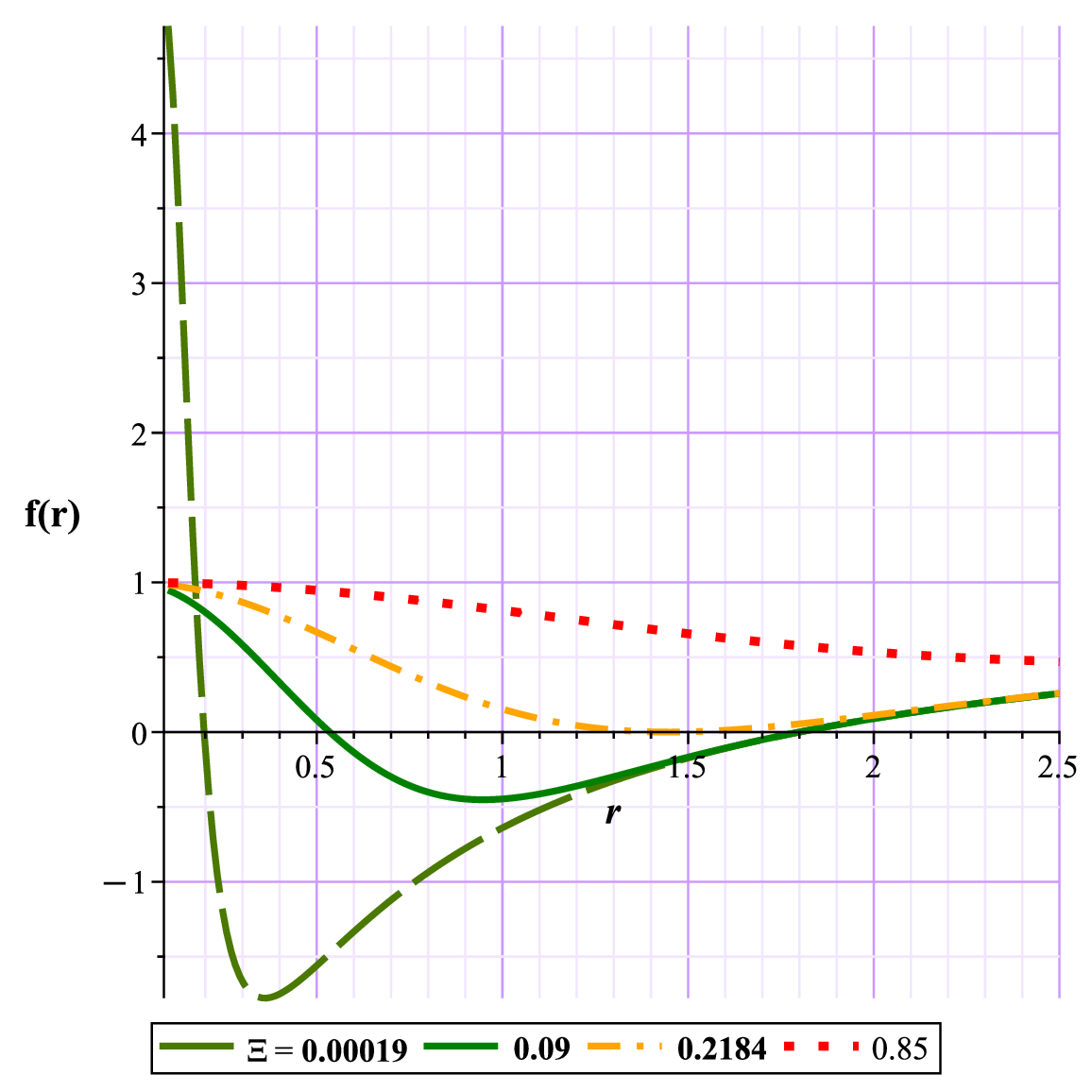}
 \label{1a}}
 \caption{\small{Metric function with different $\Xi$ for Charged NCBH  }}
 \label{1}
\end{center}
\end{figure}
\subsection{Topological Photon Sphere }
Now according to the metric function Eq. (\ref{(18)}) and also according to the following equations:
\begin{equation}\label{(20)}
f \! \left(r \right)=g \! \left(r \right),
\end{equation}
\begin{equation}\label{(21)}
h \! \left(r \right)=r,
\end{equation}

and with respect to Eq. (\ref{(7)}), Eq. (\ref{(8)}) we have:

\begin{equation*}\label{(0)}
\begin{split}
& h_{1}=\frac{{\mathrm{erf}\! \left(\frac{\sqrt{\frac{r^{2}}{\Xi}}}{2}\right)}^{2} q^{2} \pi}{2}-\left(-\frac{\sqrt{\pi}\, \sqrt{\frac{1}{\Xi}}\, \sqrt{2}\, q^{2}}{4}+\pi  m \right) r \mathrm{erf}\! \left(\frac{\sqrt{\frac{r^{2}}{\Xi}}}{2}\right),\\
& h_{2}=\sqrt{\Xi}\, \left(h_{1}+\left(\sqrt{\frac{r^{2}}{\Xi}}\, \left(-\frac{\sqrt{2}\, q^{2} \sqrt{\frac{1}{\Xi}}}{4}+m \sqrt{\pi}\right) {\mathrm e}^{-\frac{r^{2}}{4 \Xi}}+\frac{r \pi}{2}\right) r \right),\\
\end{split}
\end{equation*}
\begin{equation}\label{(22)}
\begin{split}
H =\frac{\sqrt{2}\, \sqrt{\frac{-\frac{\mathrm{erf}\left(\frac{\sqrt{2}\, \sqrt{\frac{r^{2}}{\Xi}}}{2}\right) r \sqrt{2}\, q^{2} \sqrt{\pi}}{4}+h_{2}}{\sqrt{\Xi}\, r^{2}}}\, \csc \! \left(\theta \right)}{\sqrt{\pi}\, r}.
\end{split}
\end{equation}
\begin{equation*}\label{(0)}
\begin{split}
& \varphi_{1}=\frac{\Xi^{\frac{3}{2}} \sqrt{\frac{r^{2}}{\Xi}}\, \pi  \sqrt{2}\, \mathrm{erf}\! \left(\frac{\sqrt{2}\, \sqrt{\frac{r^{2}}{\Xi}}}{2}\right) q^{2} r}{4}-\frac{2 \Xi^{2} \sqrt{\frac{r^{2}}{\Xi}}\, \pi^{\frac{3}{2}} {\mathrm{erf}\! \left(\frac{\sqrt{\frac{r^{2}}{\Xi}}}{2}\right)}^{2} q^{2}}{3},\\
& \varphi_{2}=\frac{\left(\left(-\frac{\sqrt{\pi}\, \sqrt{\frac{1}{\Xi}}\, \sqrt{2}\, q^{2}}{4}+\pi  m \right) \left(r^{2}+6 \Xi \right) r {\mathrm e}^{-\frac{r^{2}}{4 \Xi}}+\sqrt{\pi}\, {\mathrm e}^{-\frac{r^{2}}{2 \Xi}} q^{2} r \sqrt{\Xi}+2 \Xi^{2} \sqrt{\frac{r^{2}}{\Xi}}\, \pi^{\frac{3}{2}}\right) r^{2}}{6},\\
\end{split}
\end{equation*}
\begin{equation}\label{(23)}
\begin{split}
\phi^{r}=\frac{3 \csc \! \left(\theta \right) \left(\varphi_{1}+\left(\frac{{\mathrm e}^{-\frac{r^{2}}{4 \Xi}} \pi  q^{2} r}{3}+\Xi  \sqrt{\frac{r^{2}}{\Xi}}\, \left(-\frac{\pi  \sqrt{\frac{1}{\Xi}}\, \sqrt{2}\, q^{2}}{4}+\pi^{\frac{3}{2}} m \right)\right) \Xi  r \mathrm{erf}\! \left(\frac{\sqrt{\frac{r^{2}}{\Xi}}}{2}\right)-\varphi_{2}\right)}{\pi^{\frac{3}{2}} \sqrt{\frac{r^{2}}{\Xi}}\, \Xi^{2} r^{4}}.
\end{split}
\end{equation}
\begin{equation*}\label{(0)}
\begin{split}
& \phi^{1} =-\frac{\mathrm{erf}\! \left(\frac{\sqrt{2}\, \sqrt{\frac{r^{2}}{\Xi}}}{2}\right) r \sqrt{2}\, q^{2} \sqrt{\pi}}{4},\\
& \phi^{2}=\frac{{\mathrm{erf}\! \left(\frac{\sqrt{\frac{r^{2}}{\Xi}}}{2}\right)}^{2} q^{2} \pi}{2}-\left(-\frac{\sqrt{\pi}\, \sqrt{\frac{1}{\Xi}}\, \sqrt{2}\, q^{2}}{4}+\pi  m \right) r \mathrm{erf}\! \left(\frac{\sqrt{\frac{r^{2}}{\Xi}}}{2}\right),\\
\end{split}
\end{equation*}
\begin{equation}\label{(24)}
\begin{split}
\phi^{\theta}=-\frac{\sqrt{2}\, \sqrt{\frac{\phi^{1} +\sqrt{\Xi}\, \left(\phi^{2}+\left(\sqrt{\frac{r^{2}}{\Xi}}\, \left(-\frac{\sqrt{2}\, q^{2} \sqrt{\frac{1}{\Xi}}}{4}+m \sqrt{\pi}\right) {\mathrm e}^{-\frac{r^{2}}{4 \Xi}}+\frac{r \pi}{2}\right) r \right)}{\sqrt{\Xi}\, r^{2}}}\, \csc \! \left(\theta \right) \cot \! \left(\theta \right)}{\sqrt{\pi}\, r^{2}}.
\end{split}
\end{equation}
\begin{center}
\textbf{Case I: TTC =-1}
\end{center}
\begin{figure}[H]
 \begin{center}
 \subfigure[]{
 \includegraphics[height=6.5cm,width=8cm]{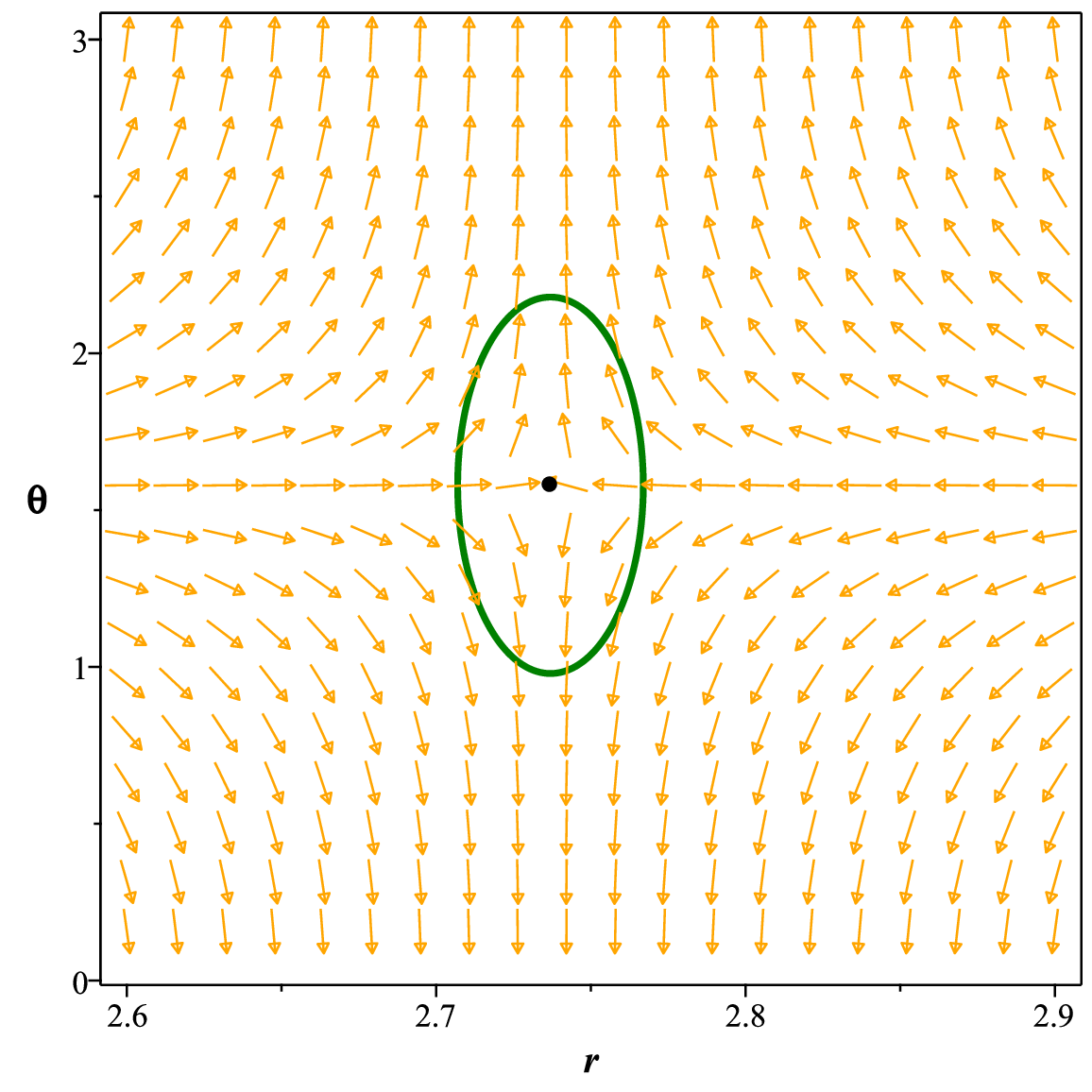}
 \label{2a}}
 \subfigure[]{
 \includegraphics[height=6.5cm,width=8cm]{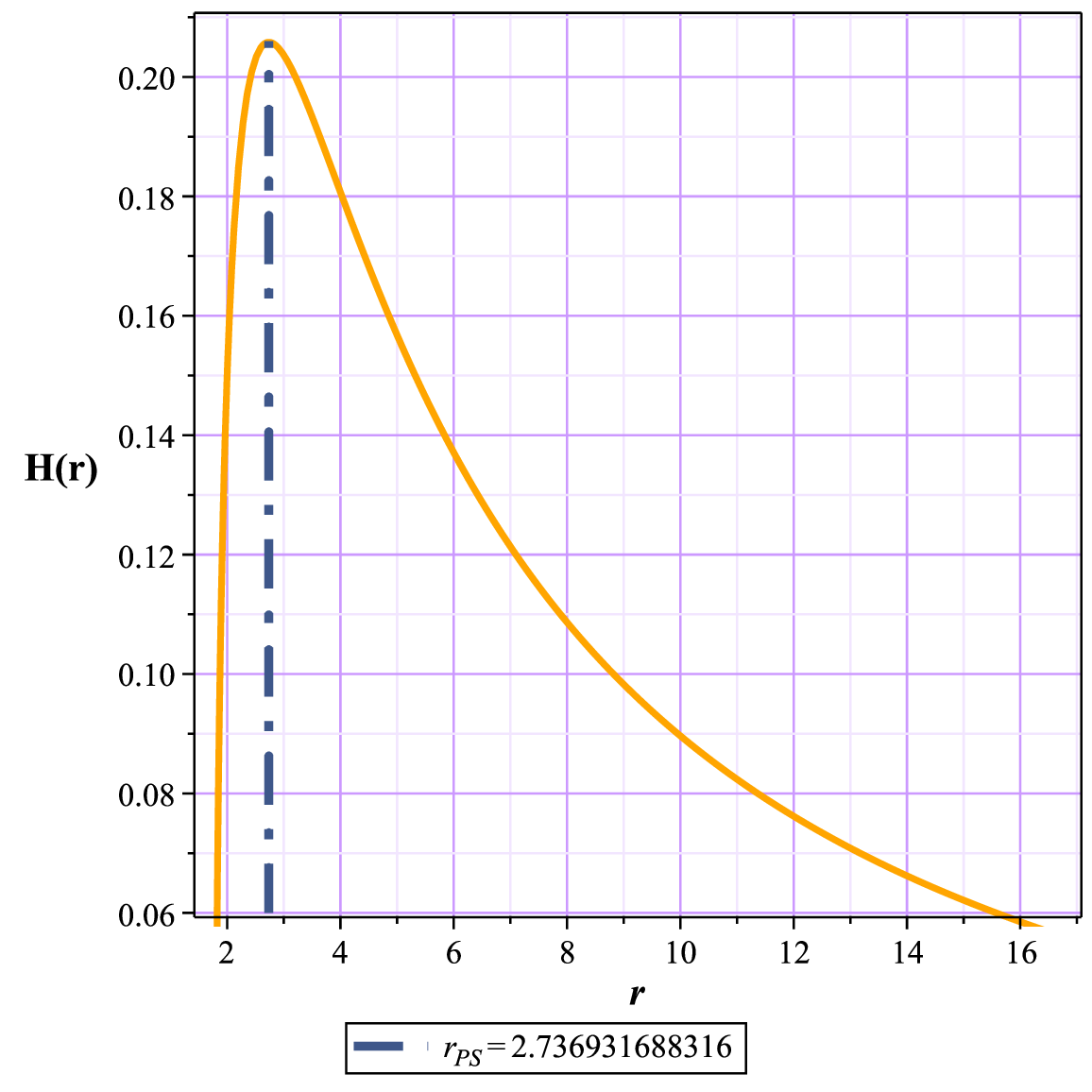}
 \label{2b}}
 
   \caption{\small{Fig (2a):The normal vector field $n$ in the $(r-\theta)$ plane. The photon sphere is located at $ (r,\theta)=(2.736931688316,1.57)$ with respect to $( \Xi = 0.001, q = 0.6, m = 1 )$  , (2b): the topological potential H(r) for Charged NCBH  }}
 \label{2}
\end{center}
\end{figure}
In the first scenario, with the chosen parameter value of $\Xi = 0.001$, we observe the emergence of a photon sphere outside the event horizon $r_h = 1.8$. This photon sphere has a topological charge (TTC) of -1 Fig. (2a) and, as shown in Fig. 2b, represents an unstable maximum. Consequently, this case presents a black hole containing an unstable photon sphere.
In the second scenario, with $\Xi = 0.31$, the gravitational structure in this horizonless region exhibits two photon spheres with topological charges of -1 and 1, indicating a spacetime with a TTC of zero Fig.(3a). From an energy perspective Fig. (3b), this situation corresponds to the presence of both a minimum and a maximum. Under these conditions, the structure manifests as a naked singularity.
\begin{center}
\textbf{Case II: TTC = 0 }
\end{center}
\begin{figure}[H]
 \begin{center}
 \subfigure[]{
 \includegraphics[height=6.5cm,width=8cm]{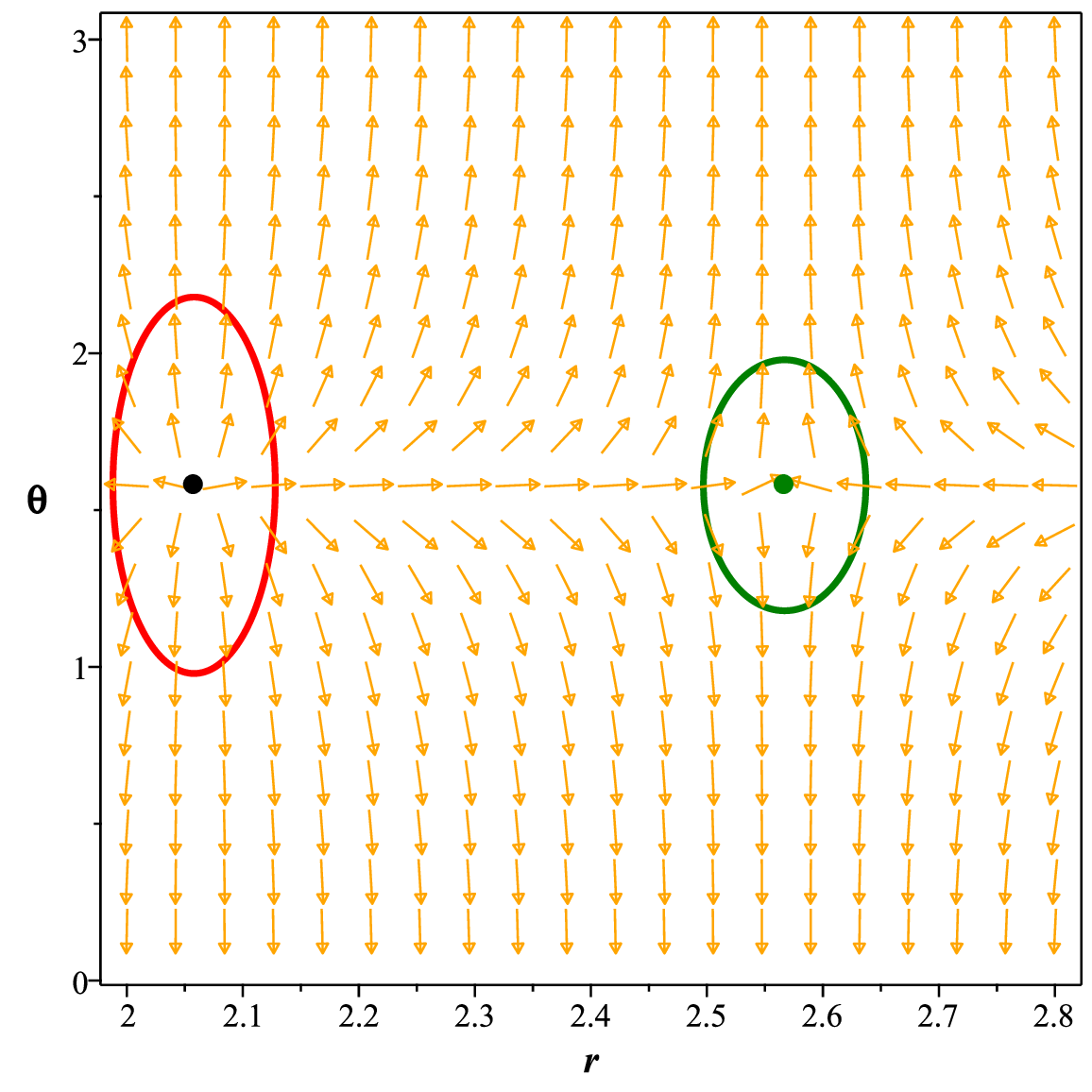}
 \label{3a}}
 \subfigure[]{
 \includegraphics[height=6.5cm,width=8cm]{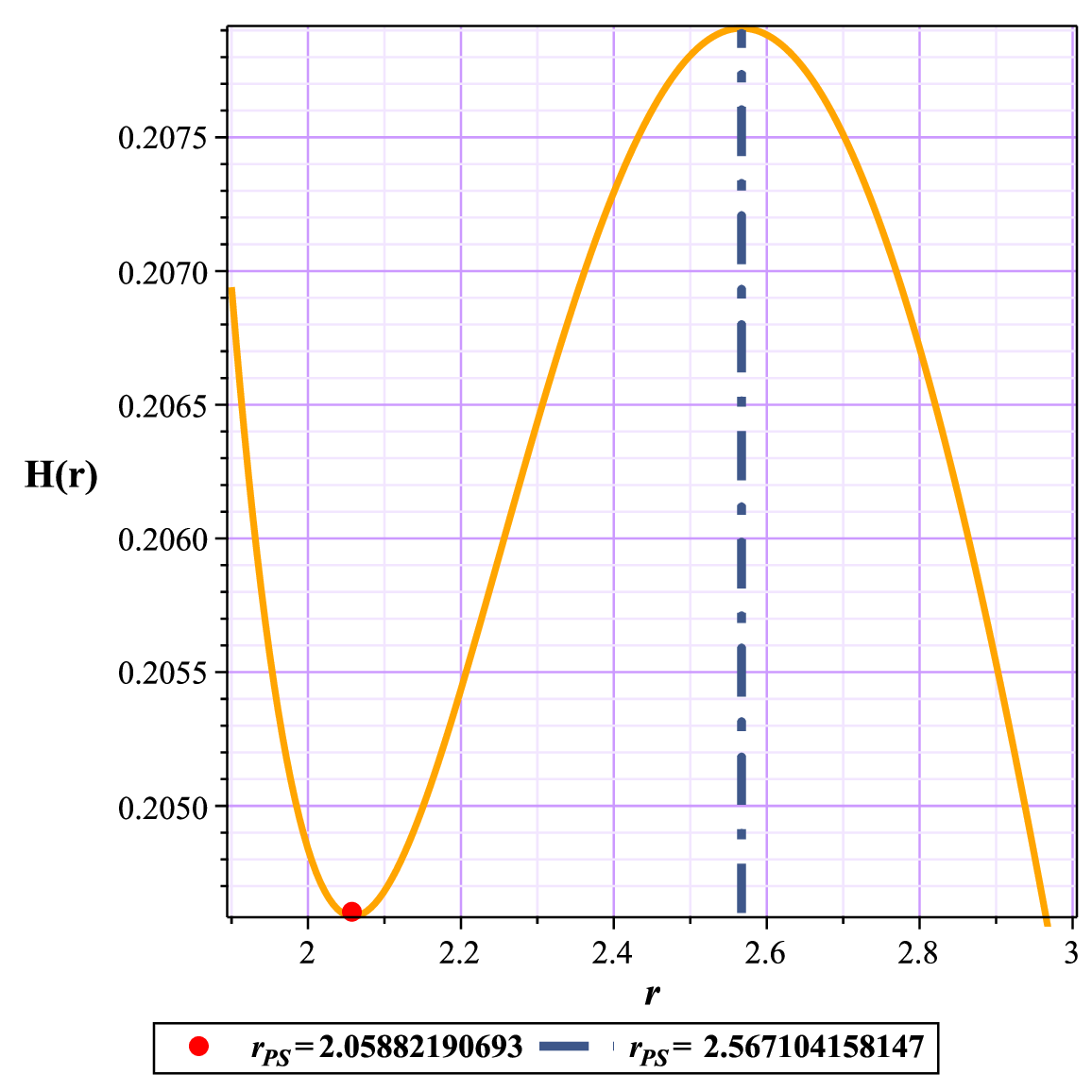}
 \label{3b}}
 
   \caption{\small{Fig (3a):The normal vector field $n$ in the $(r-\theta)$ plane. The photon spheres are located at $ (r,\theta)=(2.05882190693,1.57)$ and $ (r,\theta)=(2.567104158147,1.57)$  with respect to $( \Xi = 0.31, q = 0.6, m = 1 )$  , (3b): the topological potential H(r) for Charged NCBH  }}
 \label{3}
\end{center}
\end{figure}
Now, we must address whether the structure will exhibit a naked singularity for all values of $\Xi > 0.2184 $, or if there are parametric constraints for the naked singularity as well. Our results (Table 1) indicate that the structure will only exhibit a naked singularity up to $\Xi = 0.3273$. Beyond this value, there will be practically no physical response. It is important to note that we have focused on the range of $\Xi$ and assigned an arbitrary values to m, q. Naturally, changing these values can alter the parametric range.

\begin{center}
\begin{table}[H]
  \centering
\begin{tabular}{|p{3cm}|p{4cm}|p{5cm}|p{1.5cm}|p{2cm}|}
  \hline
  \centering{Charged NCBH}  & \centering{Fix parametes} &\centering{Conditions}& *TTC&\ $(R_{PLPS})$\\[3mm]
   \hline
  \centering{unstable photon sphere} & \centering $ q = 0.6, m = 1  $ & \centering{$0<\Xi \leq 0.2184 $} & $-1$&\ $2.724363554$\\[3mm]
   \hline
 \centering{naked singularity} & \centering $q = 0.6, m = 1 $ & \centering{$0.2184<\Xi \leq 0.3273 $} &\centering $0$&\ $-$ \\[3mm]
   \hline
   \centering{*Unauthorized area} & \centering $q = 0.6, m = 1 $ & \centering{$\Xi> 0.3273$} & \centering $ nothing $ &\ $-$ \\[3mm]
   \hline
   \end{tabular}
   \caption{*Unauthorized region: is the region where the roots of $\phi$ equations become negative or imaginary.\\ $R_{PLPS}$: the minimum or maximum possible radius for the appearance of an unstable photon sphere.}\label{1}
\end{table}
 \end{center}
\subsection{TCOs}
According to Eq. (\ref{(9)}) and Eq. (\ref{(10)}) and Eq. (\ref{(14)}) for this model we will have:
\begin{equation*}\label{(0)}
a_{1}=\left(-\frac{\pi  \sqrt{\frac{1}{\Xi}}\, \sqrt{2}\, q^{2}}{4}+\pi^{\frac{3}{2}} m \right) r \mathrm{erf}\! \left(\frac{\sqrt{\frac{r^{2}}{\Xi}}}{2}\right)+\sqrt{\frac{r^{2}}{\Xi}}\, \left(-\frac{\sqrt{\pi}\, \sqrt{\frac{1}{\Xi}}\, \sqrt{2}\, q^{2}}{4}+\pi  m \right) {\mathrm e}^{-\frac{r^{2}}{4 \Xi}}+\frac{\pi^{\frac{3}{2}} r}{2},
\end{equation*}
\begin{equation}\label{(25)}
A =r^{2} E^{2}-\frac{2 L^{2} \left(-\frac{\pi  \mathrm{erf}\left(\frac{\sqrt{2}\, \sqrt{\frac{r^{2}}{\Xi}}}{2}\right) \sqrt{2}\, q^{2} r}{4}+a_{1} \sqrt{\Xi}\right)}{\sqrt{\Xi}\, \pi^{\frac{3}{2}} r^{2}}.
\end{equation}

\begin{equation*}\label{(0)}
b_{1}=\frac{\pi^{\frac{3}{2}} {\mathrm{erf}\! \left(\frac{\sqrt{\frac{r^{2}}{\Xi}}}{2}\right)}^{2} q^{2}}{2}-\left(-\frac{\pi  \sqrt{\frac{1}{\Xi}}\, \sqrt{2}\, q^{2}}{4}+\pi^{\frac{3}{2}} m \right) r \mathrm{erf}\! \left(\frac{\sqrt{\frac{r^{2}}{\Xi}}}{2}\right)+\left(\sqrt{\frac{r^{2}}{\Xi}}\, \left(-\frac{\sqrt{\pi}\, \sqrt{\frac{1}{\Xi}}\, \sqrt{2}\, q^{2}}{4}+\pi  m \right) {\mathrm e}^{-\frac{r^{2}}{4 \Xi}}+\frac{\pi^{\frac{3}{2}} r}{2}\right) r, 
\end{equation*}
\begin{equation}\label{(26)}
B =\frac{2 \left(-\frac{\pi  \mathrm{erf}\left(\frac{\sqrt{2}\, \sqrt{\frac{r^{2}}{\Xi}}}{2}\right) \sqrt{2}\, q^{2} r}{4}+b_{1} \sqrt{\Xi}\right)}{\sqrt{\Xi}\, \pi^{\frac{3}{2}}}.
\end{equation}

\begin{equation*}\label{(0)}
\beta_{1}=-\frac{3 \mathrm{erf}\! \left(\frac{\sqrt{2}\, \sqrt{\frac{r^{2}}{\Xi}}}{2}\right) \pi  \sqrt{2}\, \Xi^{2} q^{2} r \sqrt{\frac{r^{2}}{\Xi}}}{2}+4 {\mathrm{erf}\! \left(\frac{\sqrt{\frac{r^{2}}{\Xi}}}{2}\right)}^{2} \pi^{\frac{3}{2}} \Xi^{\frac{5}{2}} q^{2} \sqrt{\frac{r^{2}}{\Xi}},
\end{equation*}
\begin{equation*}\label{(0)}
\beta_{2}=\beta_{1}-6 r \left(\frac{{\mathrm e}^{-\frac{r^{2}}{4 \Xi}} \pi  \,\Xi^{\frac{3}{2}} q^{2} r}{3}+\Xi^{\frac{5}{2}} \sqrt{\frac{r^{2}}{\Xi}}\, \left(-\frac{\pi  \sqrt{\frac{1}{\Xi}}\, \sqrt{2}\, q^{2}}{4}+\pi^{\frac{3}{2}} m \right)\right) \mathrm{erf}\! \left(\frac{\sqrt{\frac{r^{2}}{\Xi}}}{2}\right),
\end{equation*}

\begin{equation}\label{(27)}
\beta =\frac{\beta_{2}+\left(r \left(-\frac{\sqrt{\pi}\, \sqrt{\frac{1}{\Xi}}\, \sqrt{2}\, q^{2}}{4}+\pi  m \right) \left(\sqrt{\Xi}\, r^{2}+6 \Xi^{\frac{3}{2}}\right) {\mathrm e}^{-\frac{r^{2}}{4 \Xi}}+{\mathrm e}^{-\frac{r^{2}}{2 \Xi}} \Xi  q^{2} r \sqrt{\pi}+2 \pi^{\frac{3}{2}} \Xi^{\frac{5}{2}} \sqrt{\frac{r^{2}}{\Xi}}\right) r^{2}}{2 \pi^{\frac{3}{2}} \Xi^{\frac{5}{2}} \sqrt{\frac{r^{2}}{\Xi}}\, r^{2}},
\end{equation}
where erf is The Error Function.In this article, we examine TCOs in two distinct scenarios. The first scenario considers the structure in the form of a black hole, while the second scenario explores the structure in the absence of an event horizon, manifesting as a naked singularity.
\begin{figure}[H]
 \begin{center}
 \subfigure[]{
 \includegraphics[height=6.5cm,width=8cm]{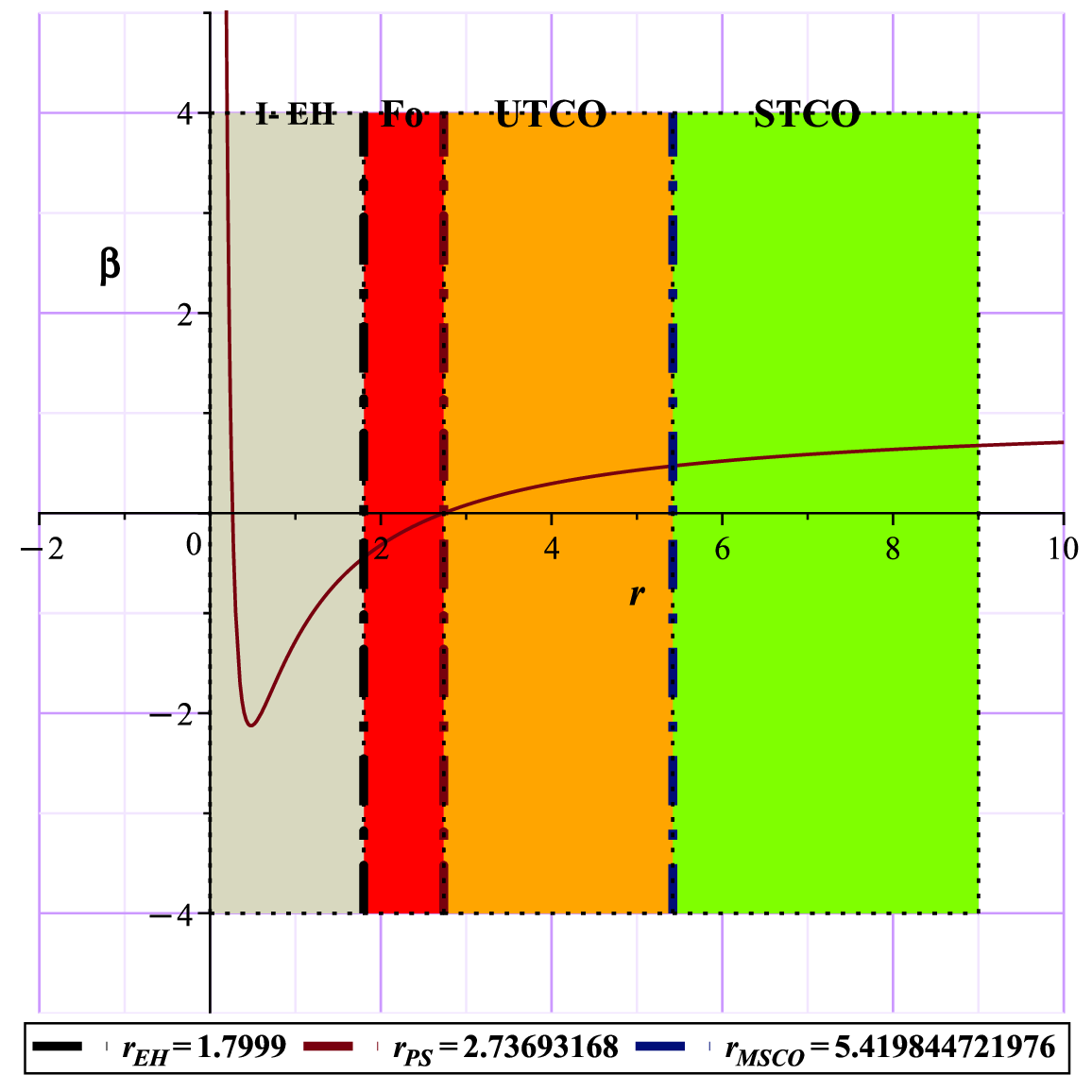}
 \label{4a}}
 \subfigure[]{
 \includegraphics[height=6.5cm,width=8cm]{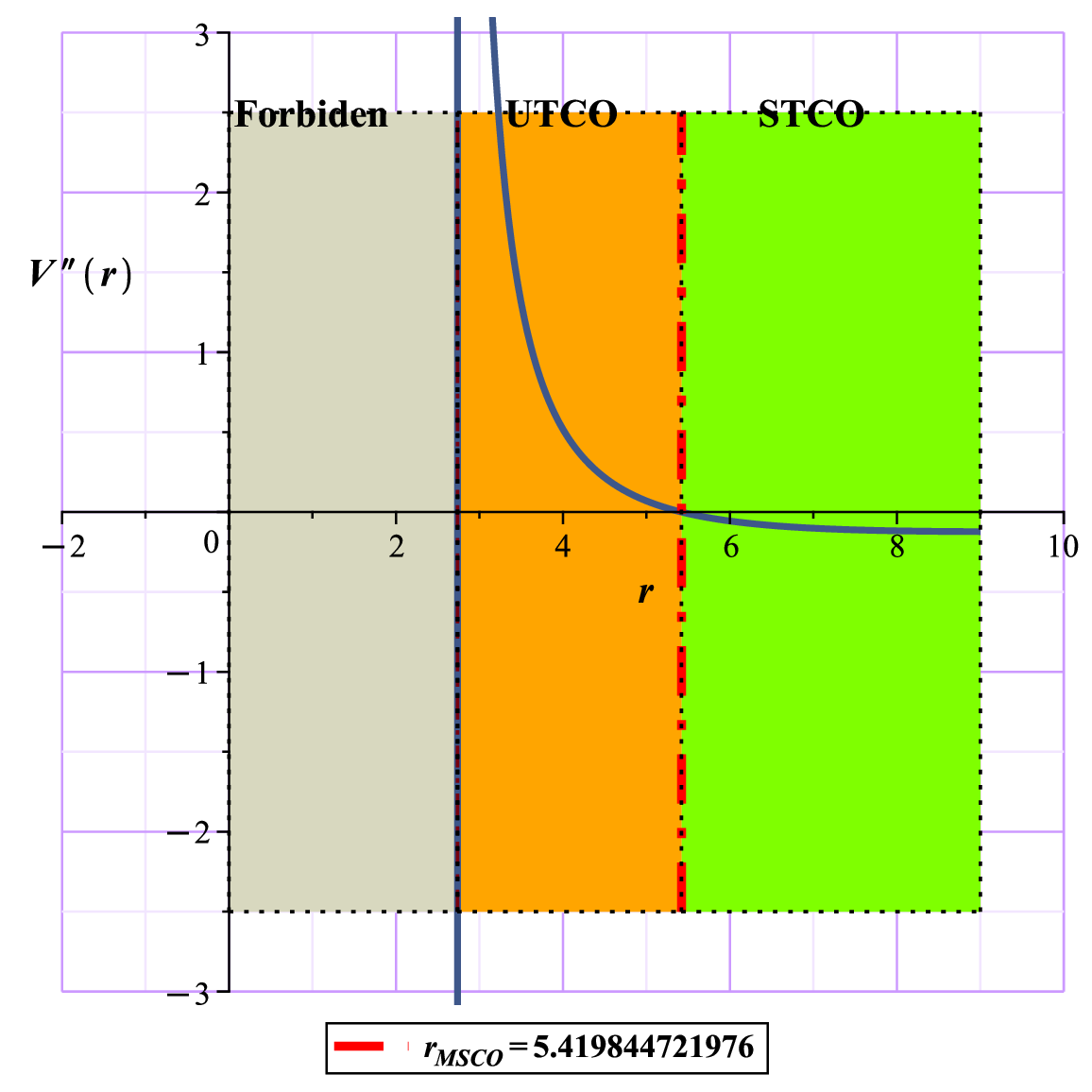}
 \label{4b}}
 
   \caption{\small{Fig (4a):$\beta$ diagram for structure in black hole form, (4b): MSCO localization and space classification for the black hole mode}}
 \label{2}
\end{center}
\end{figure}
As evident from Fig.(4a), the $\beta$ diagram shows a region with negative $\beta$ values between the event horizon and the photon sphere. Due to the negative energy, this region is considered forbidden for TCOs. In Fig.(4b), beyond the photon sphere, the ranges of TCO orbits and the boundary orbit of MSCO are also depicted.\\Now I will go to the situation where the structure does not have an event horizon.In this scenario, as observed in the previous section, two photon spheres appear in the studied spacetime. As indicated by the beta diagram in Fig.(5a), the region between these two photon spheres is forbidden for TCOs due to the negative energy. An interesting aspect of this scenario is the possibility of STCO/UTCOs existing beyond the stable photon sphere ,closer to the naked core, and two distinct boundaries(MSCO) on either side of the photon sphere, as shown in Fig.(5b).
\begin{figure}[H]
 \begin{center}
 \subfigure[]{
 \includegraphics[height=6.5cm,width=8cm]{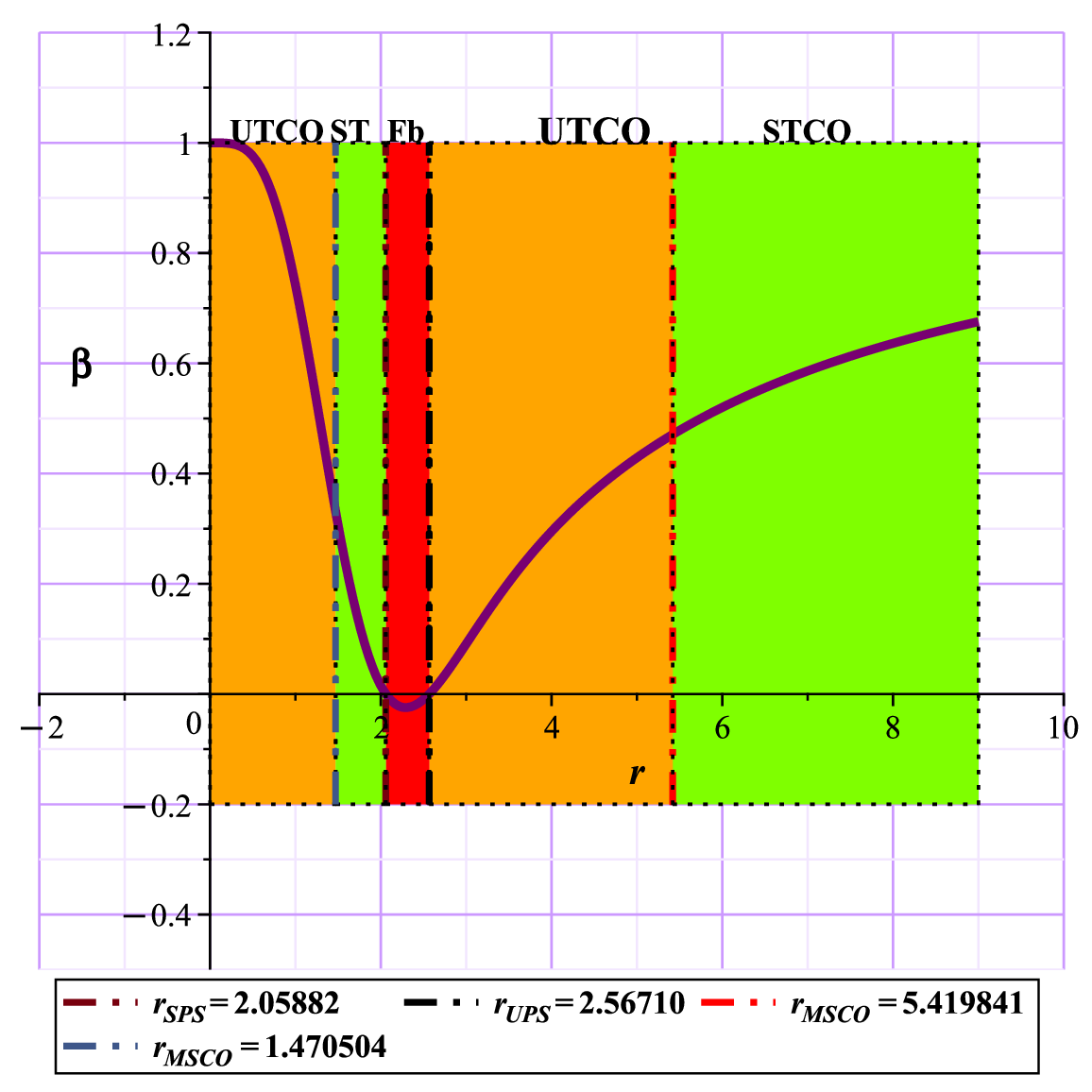}
 \label{5a}}
 \subfigure[]{
 \includegraphics[height=6.5cm,width=8cm]{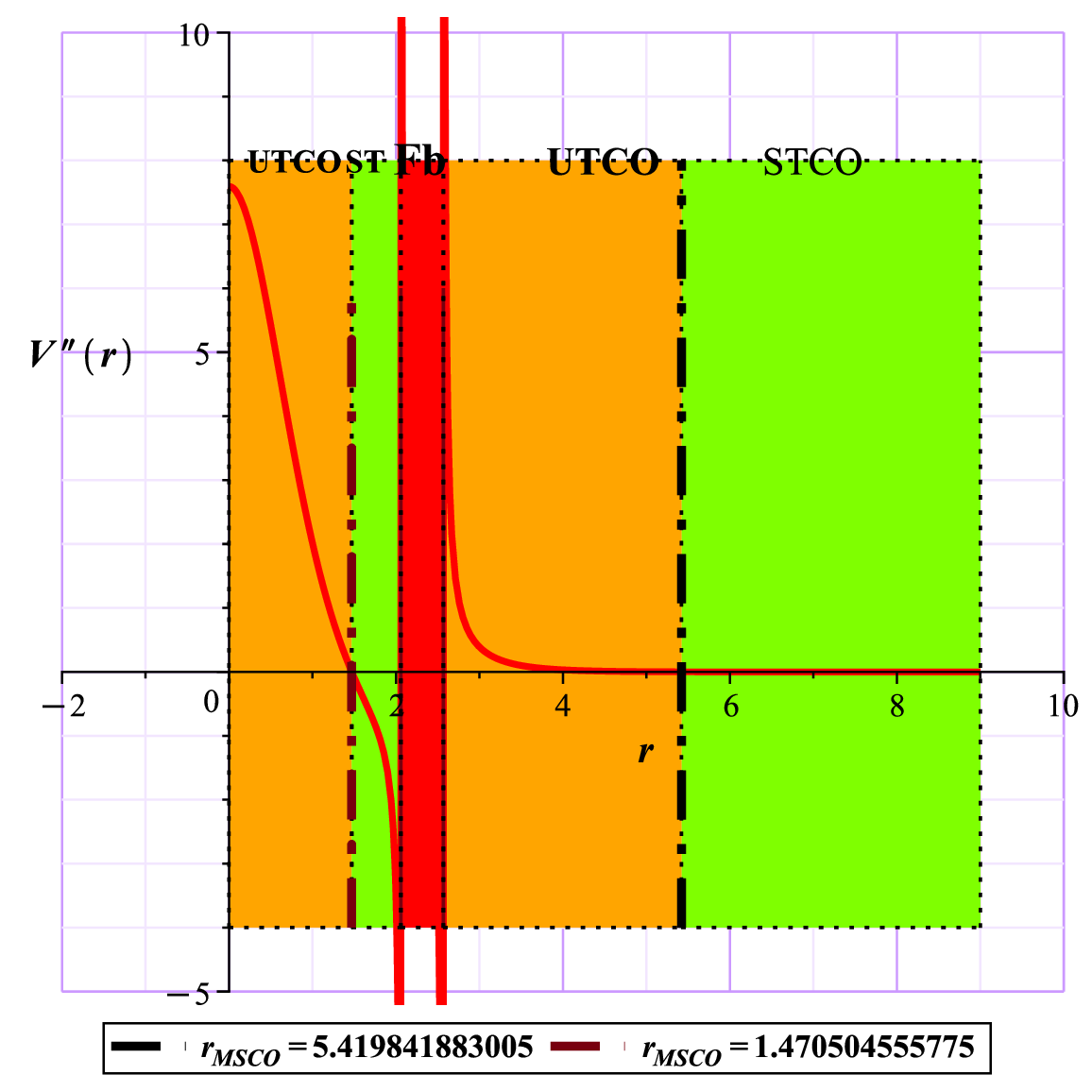}
 \label{5b}}
 
   \caption{\small{Fig (5a):$\beta$ diagram for structure in naked singularity form, (5b): MSCOs localization and space classification for the naked singularity mode}}
 \label{5}
\end{center}
\end{figure}
\section{Non-Commutative Einstein-Born-Infeld black hole}
In the previous section, we examined the mutual influence of electric charge and the NC parameter on the structure of spacetime geodesics. We now aim to investigate the impact of the field in its nonlinear form and the incorporation of constant curvature into the model.\\ The Einstein-Born-Infeld black hole(BIBH) is a solution derived from the Einstein-Born-Infeld theory, which integrates general relativity with BI nonlinear electrodynamics. The concept of BIBHs emerged as physicists explored the implications of BI electrodynamics within curved spacetime. The primary motivation behind the BI theory was to address the issue of the electron's self-energy divergence in classical electrodynamics by proposing a nonlinear electrodynamics framework. This model introduces an upper bound on the electric field, thereby avoiding the problem of infinite self-energy. The theory has experienced a resurgence of interest, particularly due to its relevance in string theory, where it appears as the leading term in the low-energy effective action of open string theory \cite{39}.\\
The metric for such black hole is \cite{40,41}:
\begin{equation}\label{(28)}
f =1-\frac{4 \gamma \! \left(\frac{3}{2},\frac{r^{2}}{4 \Xi}\right) m}{\sqrt{\pi}\, r}-\Lambda  r^{2}+\frac{2 \left(1-\sqrt{1+\frac{q^{2}}{r^{4} b^{2}}}\right) r^{2} b^{2}}{3}+\frac{4 \mathrm{hypergeom}\! \left(\left[\frac{1}{4},\frac{1}{2}\right],\left[\frac{5}{4}\right],-\frac{q^{2}}{r^{4} b^{2}}\right) q^{2}}{3 r^{2}}
\end{equation}
where $\Xi$ is NC parameter,b is the BI factor parameter, m ,q and $\Lambda$ are the total mass of the source, the total electric charge and non-zero cosmological constant. $\gamma$ is the lower incomplete Gamma function and hypergeom is a generalized hypergeometric function. Due to the great influence of the cosmological constant, we will examine the structure in two situations. 
\subsection{$\Lambda<0$ }
In this model, due to the complexity of the model, we still have to solve it numerically. If we choose $ m=1, q=0.6 and \Lambda=-1$, for the graph of the metric function with respect to different $\Xi$ will have:
\begin{figure}[H]
 \begin{center}
 \subfigure[]{
 \includegraphics[height=6.5cm,width=8cm]{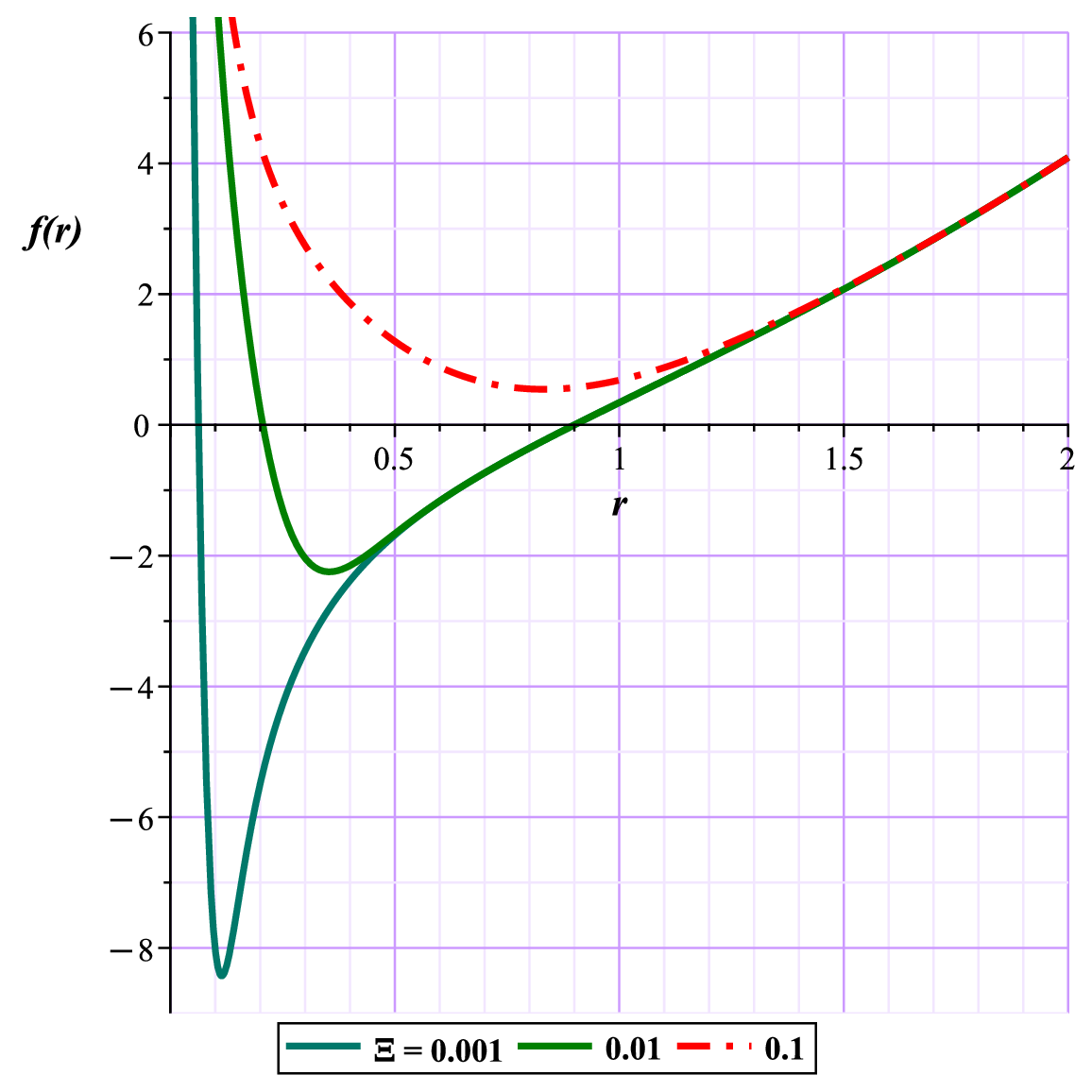}
 \label{1a}}
 \caption{\small{Metric function with different $\Xi$ for NCBIBH  }}
 \label{6}
\end{center}
\end{figure}
In this scenario, for the presence of an event horizon and the fulfillment of the WCCC, the parameter $\Xi $ is only valid up to 0.0611. This is a significant reduction compared to the previous scenario (0.2184), given the same charge and mass. The smaller the value of this parameter, the more physically realistic the model's responses become. It appears that the addition of the nonlinear BI field results in better corrections and a more realistic perspective.
\subsubsection{Topological Photon Sphere }
Now according to the metric function Eq. (\ref{(28)}) and equations Eq. (\ref{(7)}),Eq. (\ref{(8)}),Eq. (\ref{(20)}),Eq. (\ref{(21)}) we have:
\begin{equation*}\label{(0)}
h =\frac{36 \left(\frac{\sqrt{\pi}}{2}-\frac{\sqrt{\frac{r^{2}}{\Xi}}\, {\mathrm e}^{-\frac{r^{2}}{4 \Xi}}}{2}-\frac{\sqrt{\pi}\, \mathrm{erfc}\left(\frac{\sqrt{\frac{r^{2}}{\Xi}}}{2}\right)}{2}\right) m}{\sqrt{\pi}\, r}-9 \Lambda  r^{2}+6 \left(1-\sqrt{1+\frac{q^{2}}{r^{4} b^{2}}}\right) r^{2} b^{2},
\end{equation*}
\begin{equation}\label{(29)}
H =\frac{\sqrt{9+h +\frac{12 \mathrm{hypergeom}\left(\left[\frac{1}{4},\frac{1}{2}\right],\left[\frac{5}{4}\right],-\frac{q^{2}}{r^{4} b^{2}}\right) q^{2}}{r^{2}}}}{3 \sin \! \left(\theta \right) r}.
\end{equation}

\begin{equation*}\label{(0)}
\varphi_{1}=-\frac{2 \left(\frac{r^{4} b^{2}+q^{2}}{r^{4} b^{2}}\right)^{\frac{1}{4}} \pi^{\frac{3}{2}} \sqrt{2}\, \mathrm{LegendreP}\! \left(-\frac{1}{4},-\frac{1}{4},\frac{r^{4} b^{2}-q^{2}}{r^{4} b^{2}+q^{2}}\right) \Xi  b^{2} q^{2} r^{4}}{9},
\end{equation*}
\begin{equation*}\label{(0)}
\varphi_{2}=\frac{4 \sqrt{\pi}\, \mathrm{hypergeom}\! \left(\left[\frac{5}{4},\frac{3}{2}\right],\left[\frac{9}{4}\right],-\frac{q^{2}}{r^{4} b^{2}}\right) \Xi  q^{4}}{45}+r^{5} \left(\sqrt{\pi}\, \mathrm{erf}\! \left(\frac{\sqrt{\frac{r^{2}}{\Xi}}}{2}\right) \Xi  m -\frac{\sqrt{\frac{r^{2}}{\Xi}}\, m \left(r^{2}+6 \Xi \right) {\mathrm e}^{-\frac{r^{2}}{4 \Xi}}}{6}-\frac{\sqrt{\pi}\, \Xi  r}{3}\right) b^{2},
\end{equation*}
\begin{equation}\label{(30)}
\varphi_{r}=\frac{3 \csc \! \left(\theta \right) \left(\varphi_{1}+\left(-\frac{q^{2}}{r^{4} b^{2}}\right)^{\frac{1}{8}} \left(\sqrt{\frac{r^{4} b^{2}+q^{2}}{r^{4} b^{2}}}\, \varphi_{2}+\frac{2 q^{2} \Xi  \sqrt{\pi}\, r^{4} b^{2}}{9}\right) \Gamma \! \left(\frac{3}{4}\right)\right)}{\left(-\frac{q^{2}}{r^{4} b^{2}}\right)^{\frac{1}{8}} \sqrt{\pi}\, \sqrt{\frac{r^{4} b^{2}+q^{2}}{r^{4} b^{2}}}\, \Xi  r^{8} b^{2} \Gamma \! \left(\frac{3}{4}\right)}.
\end{equation}

\begin{equation*}\label{(0)}
\varphi^{1}=9-\frac{36 \left(\frac{\sqrt{\pi}}{2}-\frac{\sqrt{\frac{r^{2}}{\Xi}}\, {\mathrm e}^{-\frac{r^{2}}{4 \Xi}}}{2}-\frac{\sqrt{\pi}\, \mathrm{erfc}\left(\frac{\sqrt{\frac{r^{2}}{\Xi}}}{2}\right)}{2}\right) m}{\sqrt{\pi}\, r}-9 \Lambda  r^{2}+6 \left(1-\sqrt{1+\frac{q^{2}}{r^{4} b^{2}}}\right) r^{2} b^{2}
\end{equation*}
\begin{equation}\label{(31)}
\phi^{\theta}=-\frac{\sqrt{\varphi^{1}+\frac{12 \mathrm{hypergeom}\left(\left[\frac{1}{4},\frac{1}{2}\right],\left[\frac{5}{4}\right],-\frac{q^{2}}{r^{4} b^{2}}\right) q^{2}}{r^{2}}}\, \cos \! \left(\theta \right)}{3 \sin \! \left(\theta \right)^{2} r^{2}}
\end{equation}
\begin{center}
\textbf{TTC =-1,TTC =-1}
\end{center}
\begin{figure}[H]
 \begin{center}
 \subfigure[]{
 \includegraphics[height=5.5cm,width=8cm]{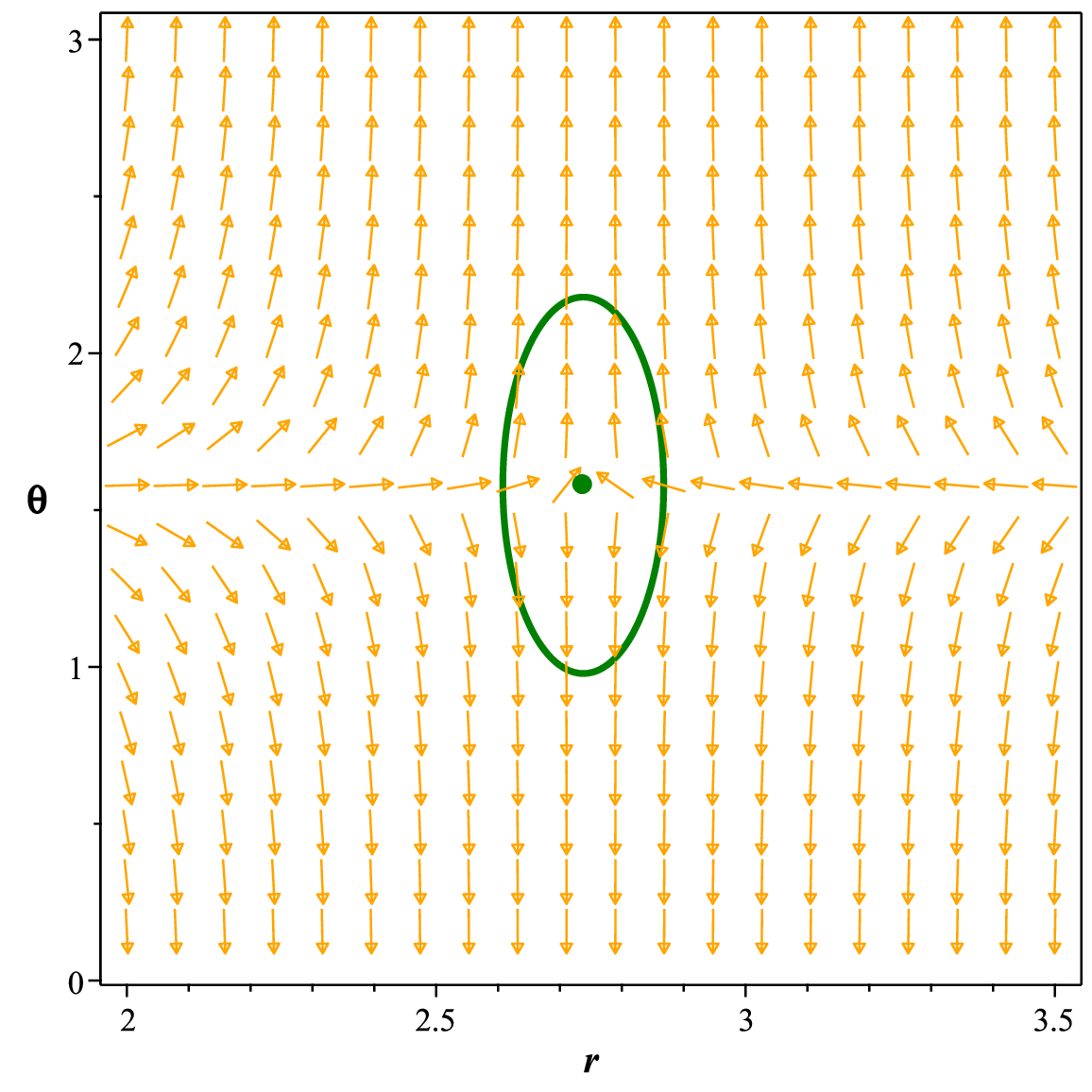}
 \label{7a}}
 \subfigure[]{
 \includegraphics[height=5.5cm,width=8cm]{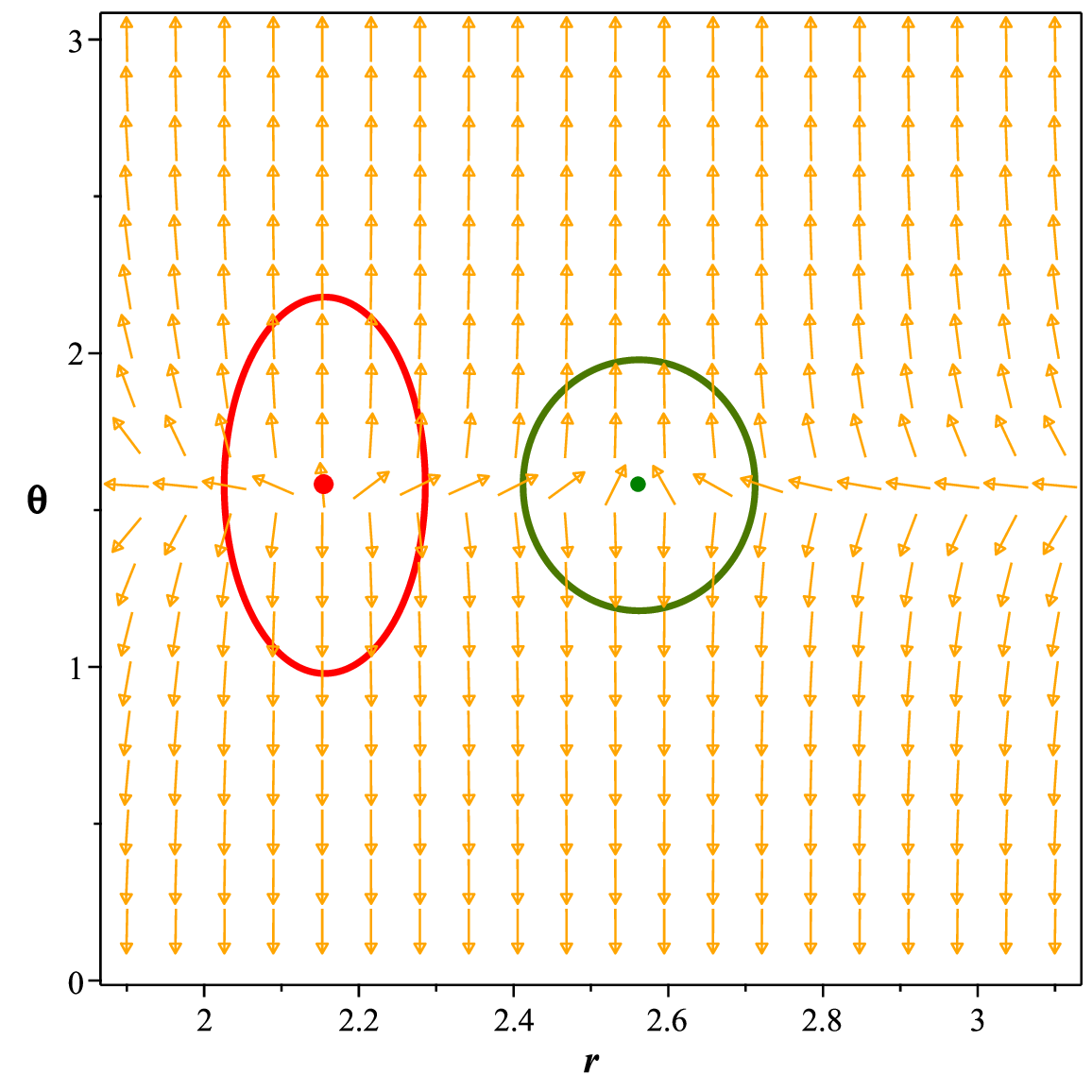}
 \label{7b}}
 
   \caption{\small{Fig (7a):The normal vector field $n$ in the $(r-\theta)$ plane. The photon sphere is located at $ (r,\theta)=(2.73763998365,1.57)$ with respect to $(\Lambda=-1,b=0.5 ,\Xi = 0.001, q = 0.6, m = 1 )$  , (7b): The photon spheres are located at $ (r,\theta)=(2.155922549434,1.57)$ and $ (r,\theta)=(2.562304621751,1.57)$ with respect to $(\Lambda=-1,b=0.5 ,\Xi = 0.3, q = 0.6, m = 1 )$ for NCBIBH  }}
 \label{2}
\end{center}
\end{figure}
The behavioral standard in this scenario remains evident. The structure exhibits a TTC of -1 in the form of a black hole Fig.(7a) and a TTC of 0 in the form of a naked singularity Fig.(7b). However, a clear comparison of the results in this scenario (Table 2) with the previous one (Table 1) reveals that, although the overall range of $\Xi$ variations for having an ultra-compact gravitational structure (black hole or naked singularity) remains constant ($0 < \Xi < 0.3$), but the division of dominance regions in the two scenarios is entirely different. This is a crucial point to consider. We have repeatedly emphasized that the smaller the parameter $\Xi$, the closer the results are to physical reality. Here, based on the results of the photon sphere study, it is evident that the structure influenced by general electromagnetism, which in this model approximately leads to Reissner-Nordström behavior, causes the model to exhibit black hole behavior over a broader range and larger values of $\Xi$. However, when the nonlinear and more precise BI field is incorporated into the model, the model clearly shows better results, and the permissible range of $\Xi$ significantly decreases.\\ This indicates that the photon sphere, in addition to its other optical characteristics, can serve as a powerful tool for evaluating the accuracy of theoretical models in alignment with empirical observations and reality, which is what we emphasize.
\begin{center}
\begin{table}[H]
  \centering
\begin{tabular}{|p{3cm}|p{4cm}|p{5cm}|p{1.5cm}|p{2cm}|}
  \hline
  \centering{NC Born-Infeld($\Lambda<0$)}  & \centering{Fix parametes} &\centering{Conditions}& *TTC&\ $(R_{PLPS})$\\[3mm]
   \hline
  \centering{unstable photon sphere} & \centering $ b=0.5,\Lambda=-1,q = 0.6, m = 1  $ & \centering{$0< \Xi \leq 0.0611 $} & $-1$&\ $2.7376747$\\[3mm]
   \hline
 \centering{naked singularity} & \centering $b=0.5,\Lambda=-1,q = 0.6, m = 1 $ & \centering{$0.0611< \Xi \leq 0.312 $} &\centering $0$&\ $-$ \\[3mm]
   \hline
   \centering{*Unauthorized area} & \centering $b=0.5,\Lambda=-1,q = 0.6, m = 1 $ & \centering{$\Xi> 0.3273$} & \centering $ nothing $ &\ $-$ \\[3mm]
   \hline
   \end{tabular}
   \caption{*Unauthorized region: is the region where the roots of $\phi$ equations become negative or imaginary.\\ $R_{PLPS}$: the minimum or maximum possible radius for the appearance of an unstable photon sphere.}\label{1}
\end{table}
 \end{center}
\subsubsection{TCOs }
According to Eq. (\ref{(9)}) and Eq. (\ref{(10)}) and Eq. (\ref{(14)}) for this model we will have:
\begin{equation*}\label{(0)}
a =1-\frac{4 \left(\frac{\sqrt{\pi}}{2}-\frac{\sqrt{\frac{r^{2}}{\Xi}}\, {\mathrm e}^{-\frac{r^{2}}{4 \Xi}}}{2}-\frac{\sqrt{\pi}\, \mathrm{erfc}\left(\frac{\sqrt{\frac{r^{2}}{\Xi}}}{2}\right)}{2}\right) m}{\sqrt{\pi}\, r}-\Lambda  r^{2},
\end{equation*}
\begin{equation}\label{(32)}
A =E^{2} r^{2}-\left(a +\frac{2 \left(1-\sqrt{1+\frac{q^{2}}{r^{4} b^{2}}}\right) r^{2} b^{2}}{3}+\frac{4 \mathrm{hypergeom}\! \left(\left[\frac{1}{4},\frac{1}{2}\right],\left[\frac{5}{4}\right],-\frac{q^{2}}{r^{4} b^{2}}\right) q^{2}}{3 r^{2}}\right) L^{2}.
\end{equation}
\begin{equation}\label{(33)}
B =r^{2} \left(a +\frac{2 \left(1-\sqrt{1+\frac{q^{2}}{r^{4} b^{2}}}\right) r^{2} b^{2}}{3}+\frac{4 \mathrm{hypergeom}\! \left(\left[\frac{1}{4},\frac{1}{2}\right],\left[\frac{5}{4}\right],-\frac{q^{2}}{r^{4} b^{2}}\right) q^{2}}{3 r^{2}}\right).
\end{equation}

\begin{equation*}\label{(0)}
\mu_{1}=-6 m \Xi  \sqrt{\pi}\, \mathrm{erf}\! \left(\frac{\sqrt{\frac{r^{2}}{\Xi}}}{2}\right)+\sqrt{\frac{r^{2}}{\Xi}}\, m \left(r^{2}+6 \Xi \right) {\mathrm e}^{-\frac{r^{2}}{4 \Xi}}+2 r \Xi  \sqrt{\pi},
\end{equation*}
\begin{equation*}\label{(0)}
\mu_{2}=3 \Gamma \! \left(\frac{3}{4}\right) \left(\left(-\frac{8 \sqrt{\pi}\, \Xi  \mathrm{hypergeom}\! \left(\left[\frac{5}{4},\frac{3}{2}\right],\left[\frac{9}{4}\right],-\frac{q^{2}}{r^{4} b^{2}}\right) q^{4}}{15}+r^{5} b^{2} \mu_{1}\right) \sqrt{\frac{r^{4} b^{2}+q^{2}}{r^{4} b^{2}}}-\frac{4 q^{2} \Xi  \sqrt{\pi}\, r^{4} b^{2}}{3}\right),
\end{equation*}
\begin{equation}\label{(34)}
\beta =\frac{2 \left(\pi^{\frac{3}{2}} \sqrt{2}\, \mathrm{LegendreP}\! \left(-\frac{1}{4},-\frac{1}{4},\frac{r^{4} b^{2}-q^{2}}{r^{4} b^{2}+q^{2}}\right) q^{2} \Xi  r^{4} \left(\frac{r^{4} b^{2}+q^{2}}{r^{4} b^{2}}\right)^{\frac{1}{4}} b^{2}+\frac{\mu_{2} \left(-\frac{q^{2}}{r^{4} b^{2}}\right)^{\frac{1}{8}}}{4}\right)}{3 \left(-\frac{q^{2}}{r^{4} b^{2}}\right)^{\frac{1}{8}} \sqrt{\pi}\, \sqrt{\frac{r^{4} b^{2}+q^{2}}{r^{4} b^{2}}}\, \Xi  r^{6} b^{2} \Gamma \! \left(\frac{3}{4}\right)}.
\end{equation}
\begin{figure}[H]
 \begin{center}
 \subfigure[]{
 \includegraphics[height=6.5cm,width=8cm]{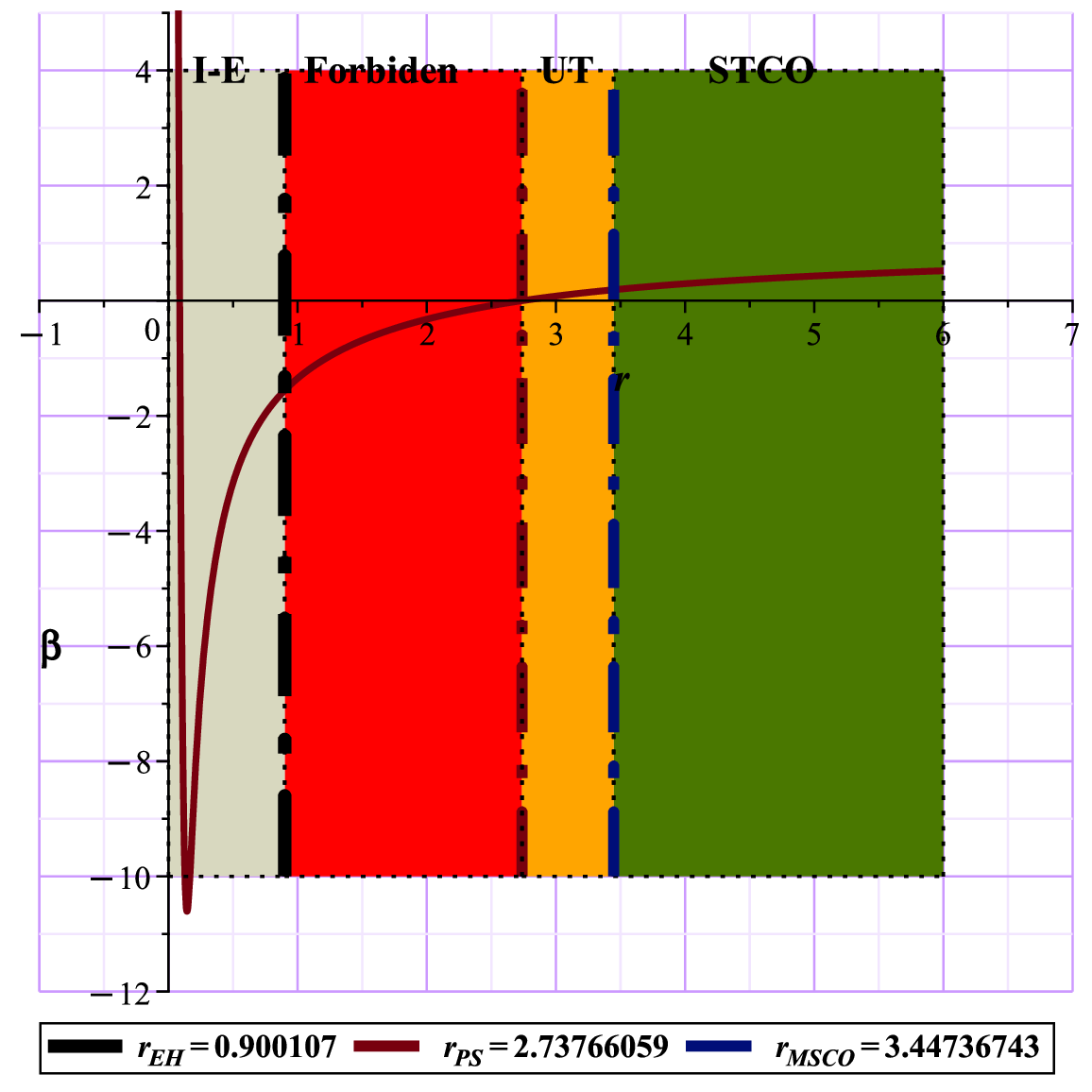}
 \label{8a}}
 \subfigure[]{
 \includegraphics[height=6.5cm,width=8cm]{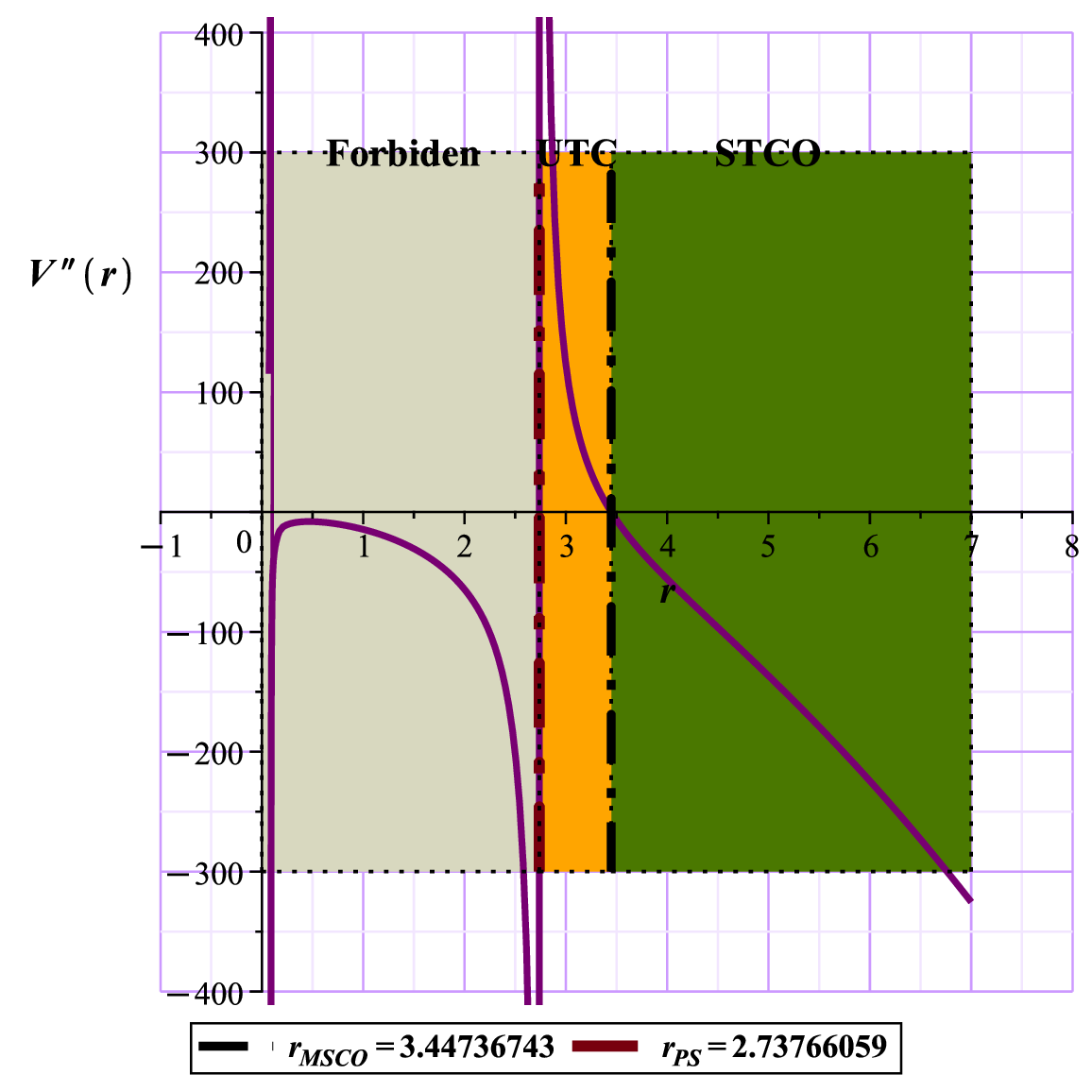}
 \label{8b}}
 
   \caption{\small{Fig (8a):$\beta$ diagram for structure in black hole form, (8b): MSCO localization and space classification for the black hole mode}}
 \label{8}
\end{center}
\end{figure}
Although the overall behavioral pattern in Fig(8) is similar to the previous scenario, a more detailed comparison reveals some differences. In Fig.(8a), the black hole region is significantly more restricted, while the forbidden region for TCOs shows a considerable increase compared to the previous scenario. Additionally, the range of UTCOs shows a noticeable decrease compared to the previous scenario, as depicted in Fig.(8b).
\begin{figure}[H]
 \begin{center}
 \subfigure[]{
 \includegraphics[height=6.5cm,width=8cm]{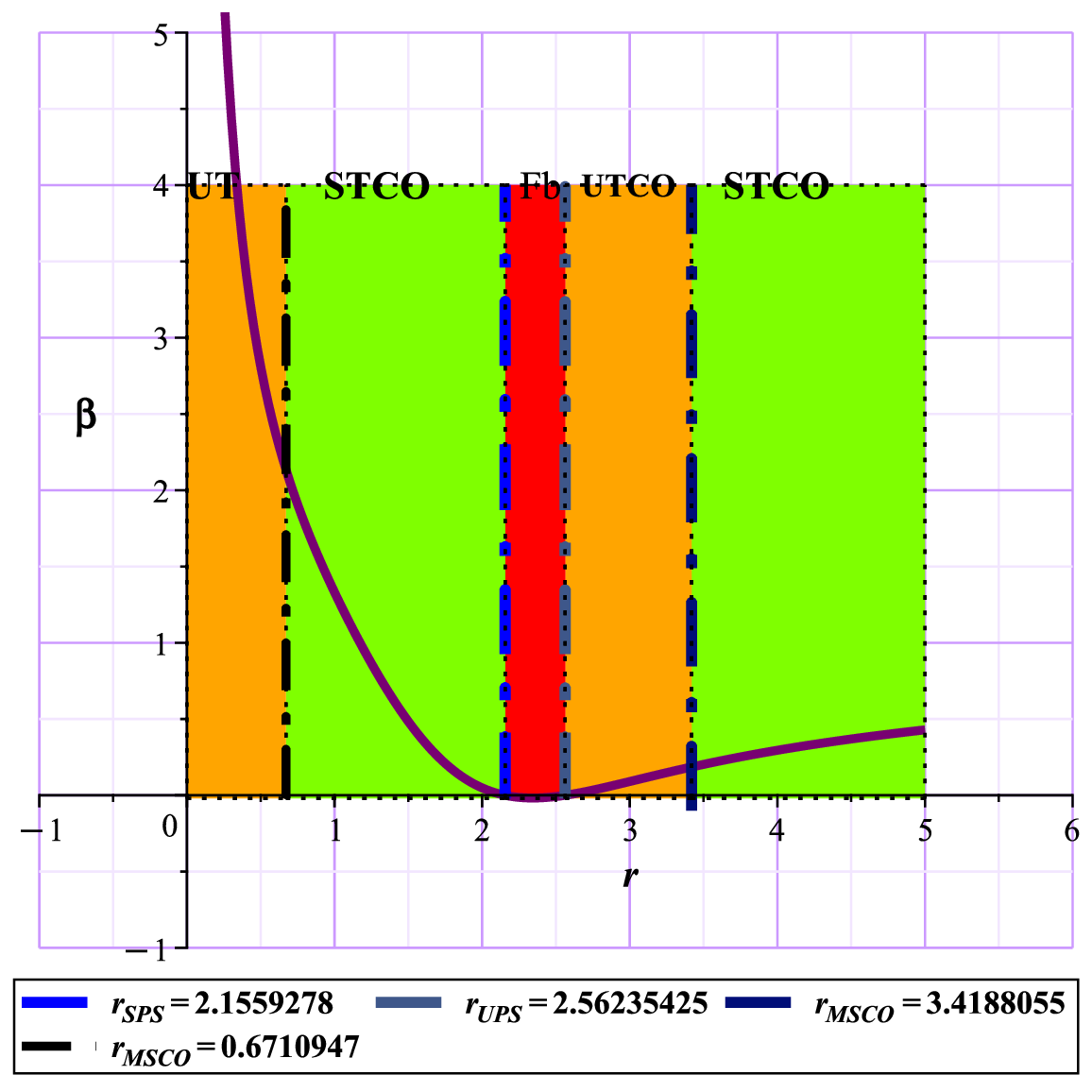}
 \label{9a}}
 \subfigure[]{
 \includegraphics[height=6.5cm,width=8cm]{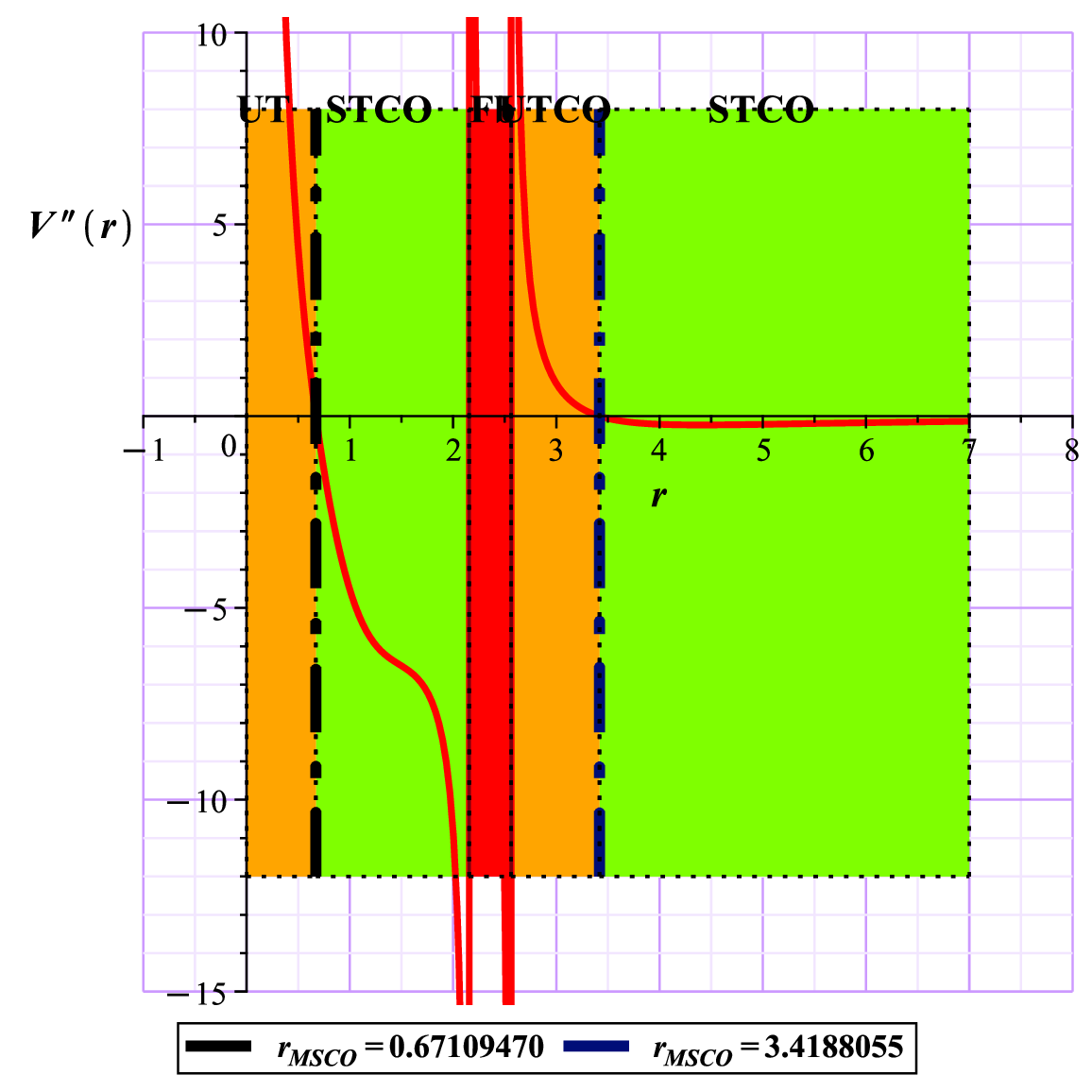}
 \label{9b}}
 
   \caption{\small{Fig (9a):$\beta$ diagram for structure in naked singularity form, (9b): MSCOs localization and space classification for the naked singularity mode}}
 \label{9}
\end{center}
\end{figure}
In the case of a naked singularity, as shown in Fig.(9), it appears that the repetitive behavioral pattern has reached complete similarity. This means that, even upon closer inspection, no significant differences can be observed compared to the previous scenario (Fig.5), whether in Fig.(9a) or Fig.(9b). Here, the presence of two boundary radii for MSCO and the existence of UTCOs near the central singularity behind the stable photon sphere can also be interesting.
\subsection{$\Lambda>0$ }
When discussing positive constant curvature or de Sitter behavior within the BI framework, it is generally stated that the system is significantly influenced by BI parameter. A critical value for this parameter can be identified, above which the black hole remains stable \cite{39}. Given this context, we aimed to determine whether the same holds true in the NC structure and identify which parameter has the most significant impact in the de Sitter scenario for the NC model. To this end, we examined the effects of the parameters b, $ \Xi $, and $ \Lambda$ in the Subextremal state (considering the mass-to-charge ratio). But first, It is important to note that in de Sitter structures, two horizons typically appear: a smaller radius horizon (analogous to the Cauchy horizon in AdS structures), which acts as the event horizon, and a significantly larger radius horizon known as the cosmological horizon. Often In the Subextremal form, the structure does not achieve the necessary stability until the cosmological horizon appears, although conditions may differ in extremal and super-extremal states\cite{40,41}.\\However, it seems that in our studied model, the presence of the cosmological horizon seems necessary since, For instance, in the first scenario, when we consider $\Lambda = +1$  our studies indicate that reducing $\Xi$ and varying  b to any desired value will not result in the emergence of the cosmological horizon,as shown in Figs.(10a) and (11a). 
\begin{figure}[H]
 \begin{center}
 \subfigure[]{
 \includegraphics[height=5.5cm,width=8cm]{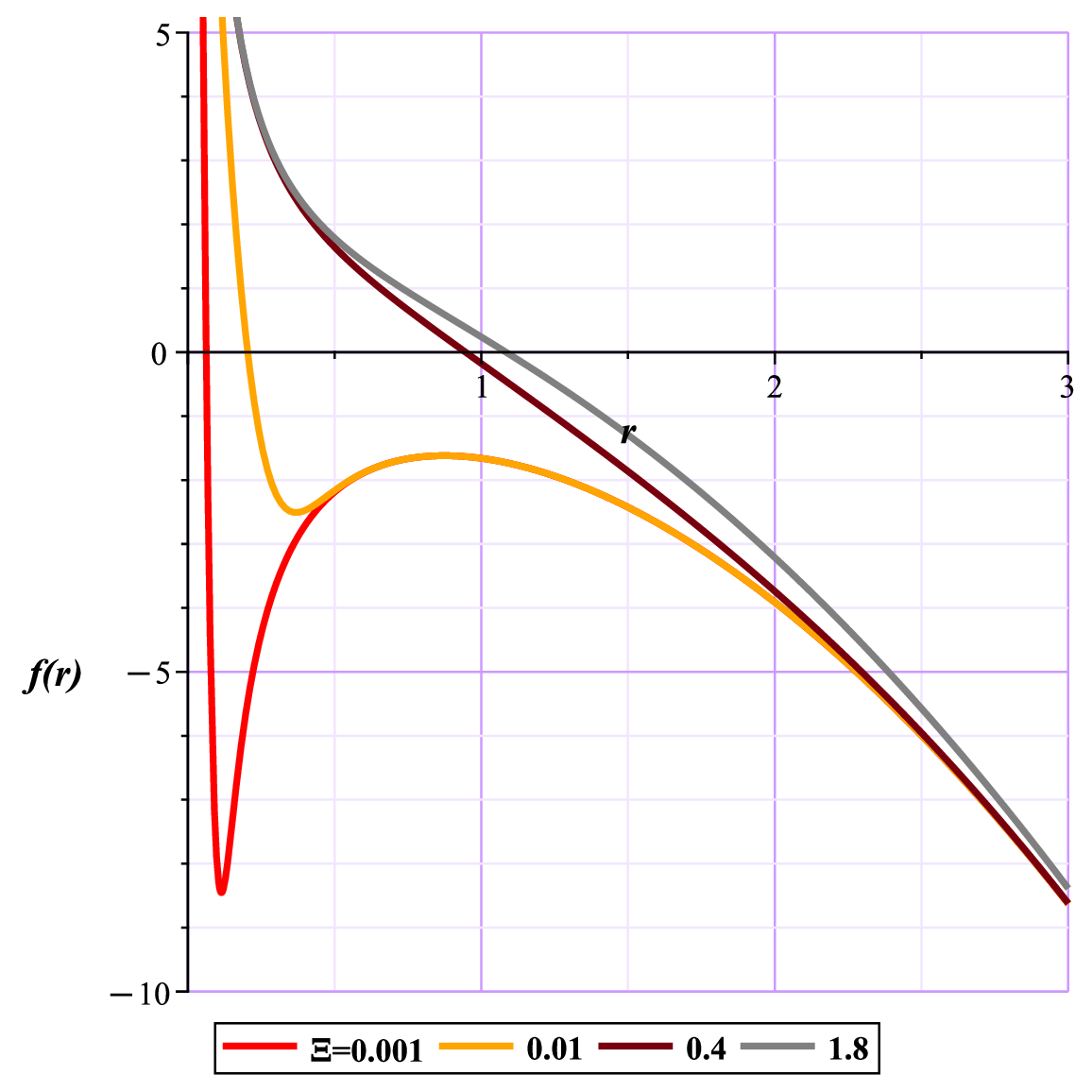}
 \label{10a}}
 \subfigure[]{
 \includegraphics[height=5.5cm,width=8cm]{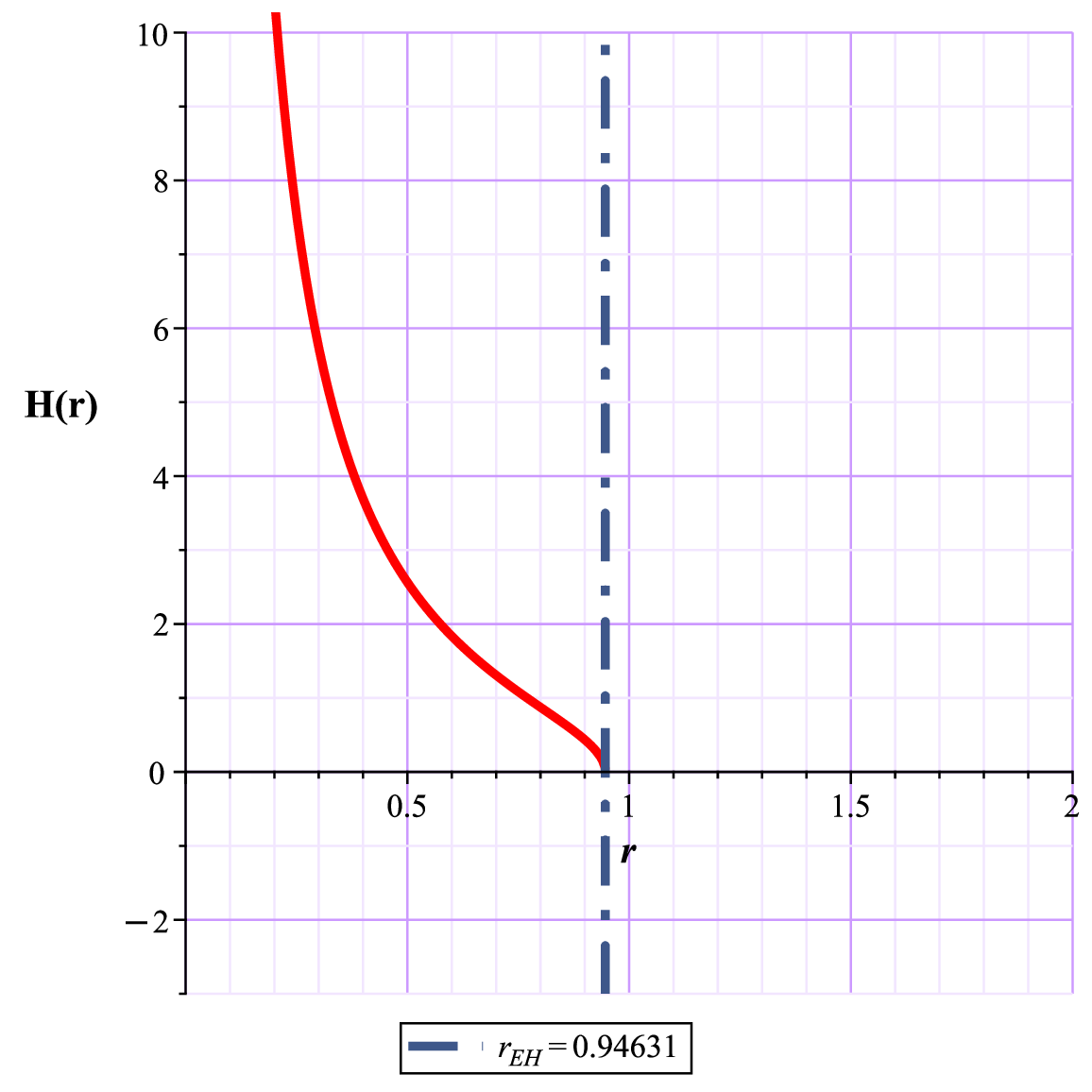}
 \label{10b}}
 
   \caption{\small{Fig (10a):Metric function with different $\Xi$ for NCBIBH with respect to $(\Lambda=+1, b=0.5,q = 0.6, m = 1 )$  , (10b): the topological potential H(r) with respect to $(\Xi=0.4,\Lambda=+1, b=0.5,q = 0.6, m = 1 )$  for NCBIBH  }}
 \label{2}
\end{center}
\end{figure}
It is noteworthy that the effective potential function of the photon sphere clearly indicates that without a  cosmic horizon, the potential function either does not extend beyond the event horizon or, even if it does extend beyond, on the one hand it loses its continuity throughout the entire of spacetime and on the other hand, this discontinuous  extension occurs in a region of space-time that is significantly distant from the probable location of the  photon sphere. Consequently, the structure will practically lack a photon sphere, as illustrated in Figs.(10b) and (11b). \textit{This aspect can itself be a positive advantage for studying the effective potential and photon sphere in dS models, as this test can serve as a preliminary indicator of the model's status and stability.}
\begin{figure}[H]
 \begin{center}
 \subfigure[]{
 \includegraphics[height=5.5cm,width=8cm]{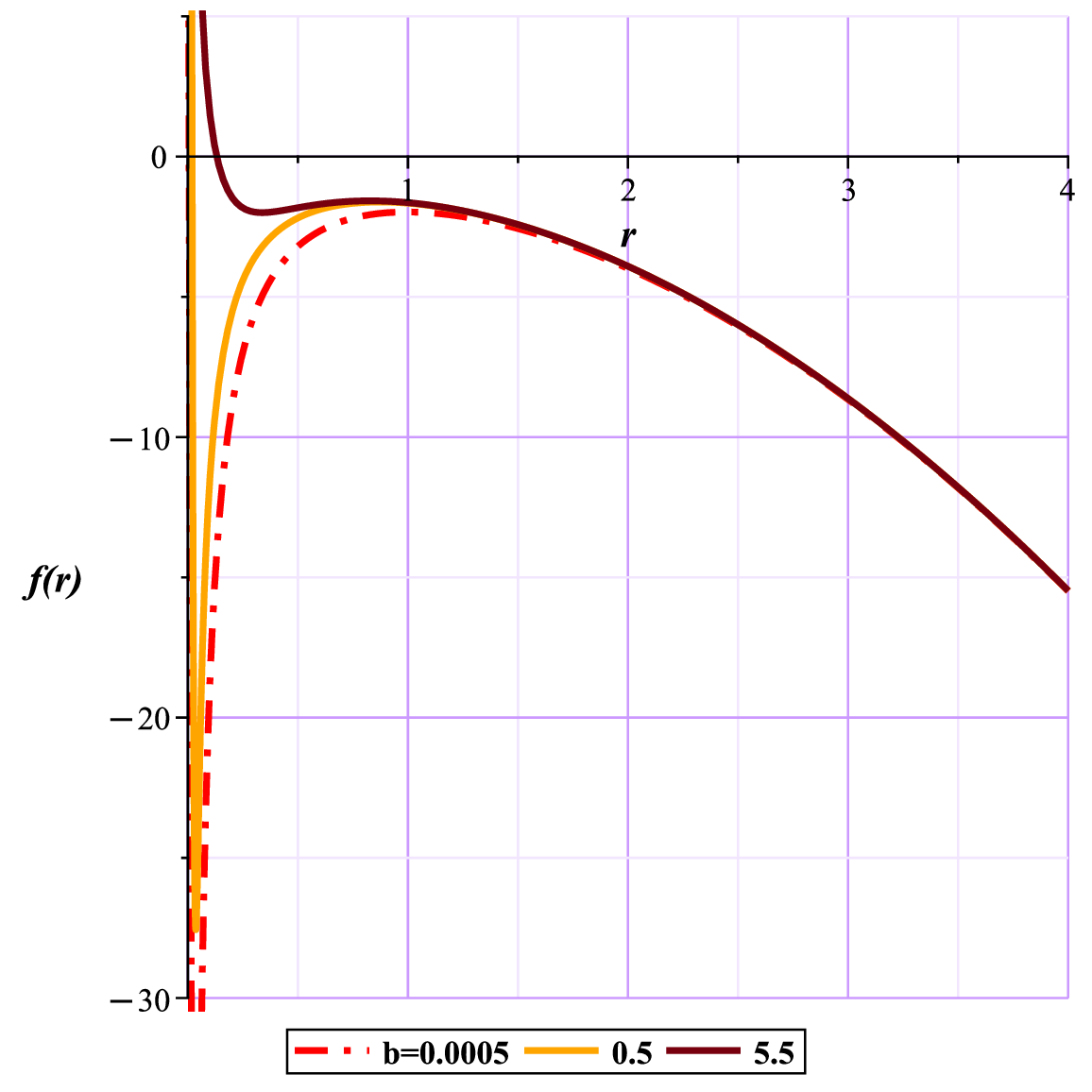}
 \label{10a}}
 \subfigure[]{
 \includegraphics[height=5.5cm,width=8cm]{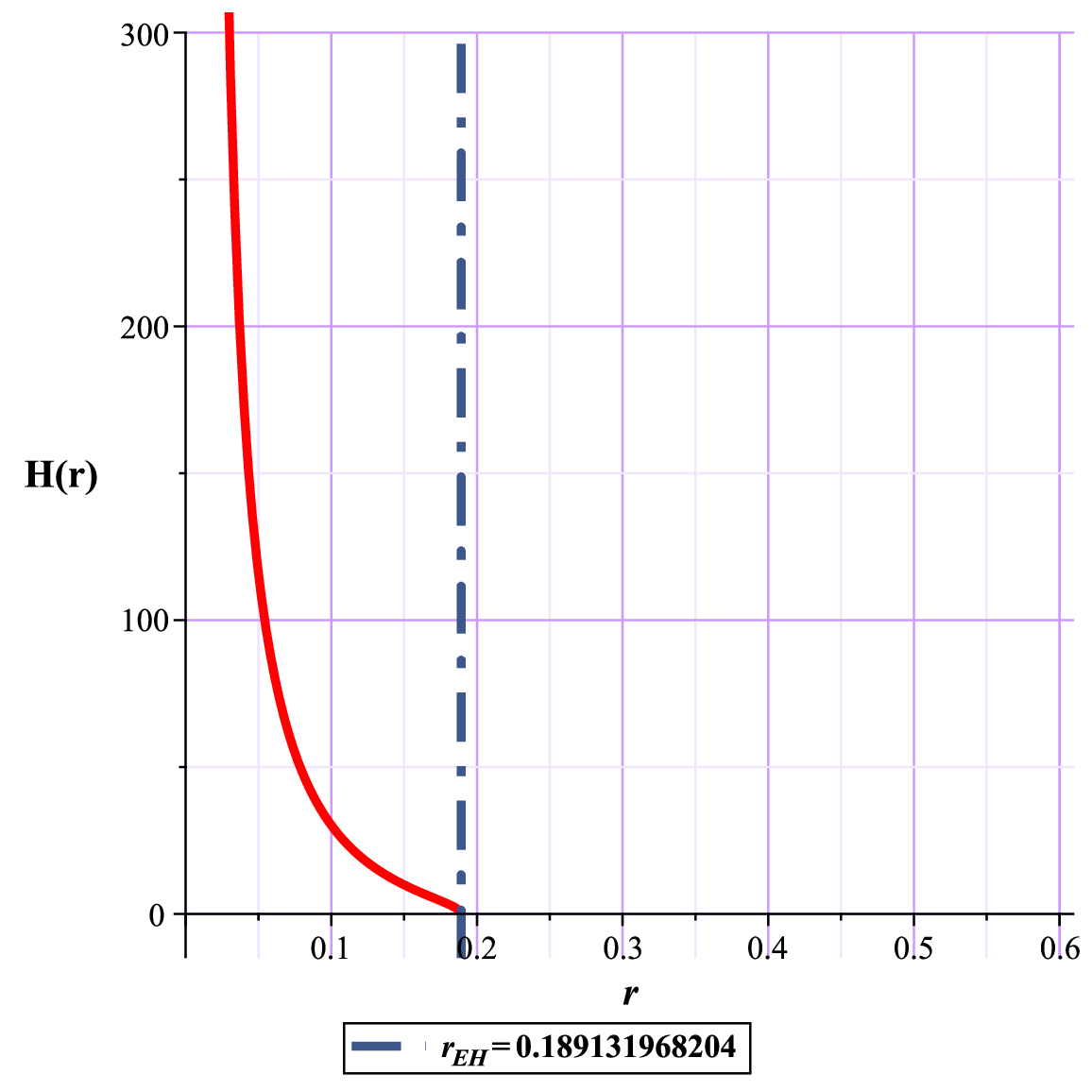}
 \label{10b}}
 
   \caption{\small{Fig (11a):Metric function with different "b" for noncommutative Born-Infeld black hole with respect to $(\Lambda=+1, \Xi=0.0001,q = 0.6, m = 1 )$  , (11b): the topological potential H(r) with respect to $(\Xi=0.0001,\Lambda=+1, b=15.5,q = 0.6, m = 1 )$  for NCBIBH  }}
 \label{2}
\end{center}
\end{figure}
Given that the model appears to be most influenced by the positive curvature constant, it is possible to classify the space based on the emergence of the photon sphere relative to a critical cosmological constant, Fig(12).

\begin{figure}[H]
 \begin{center}
 \subfigure[]{
 \includegraphics[height=5.5cm,width=8cm]{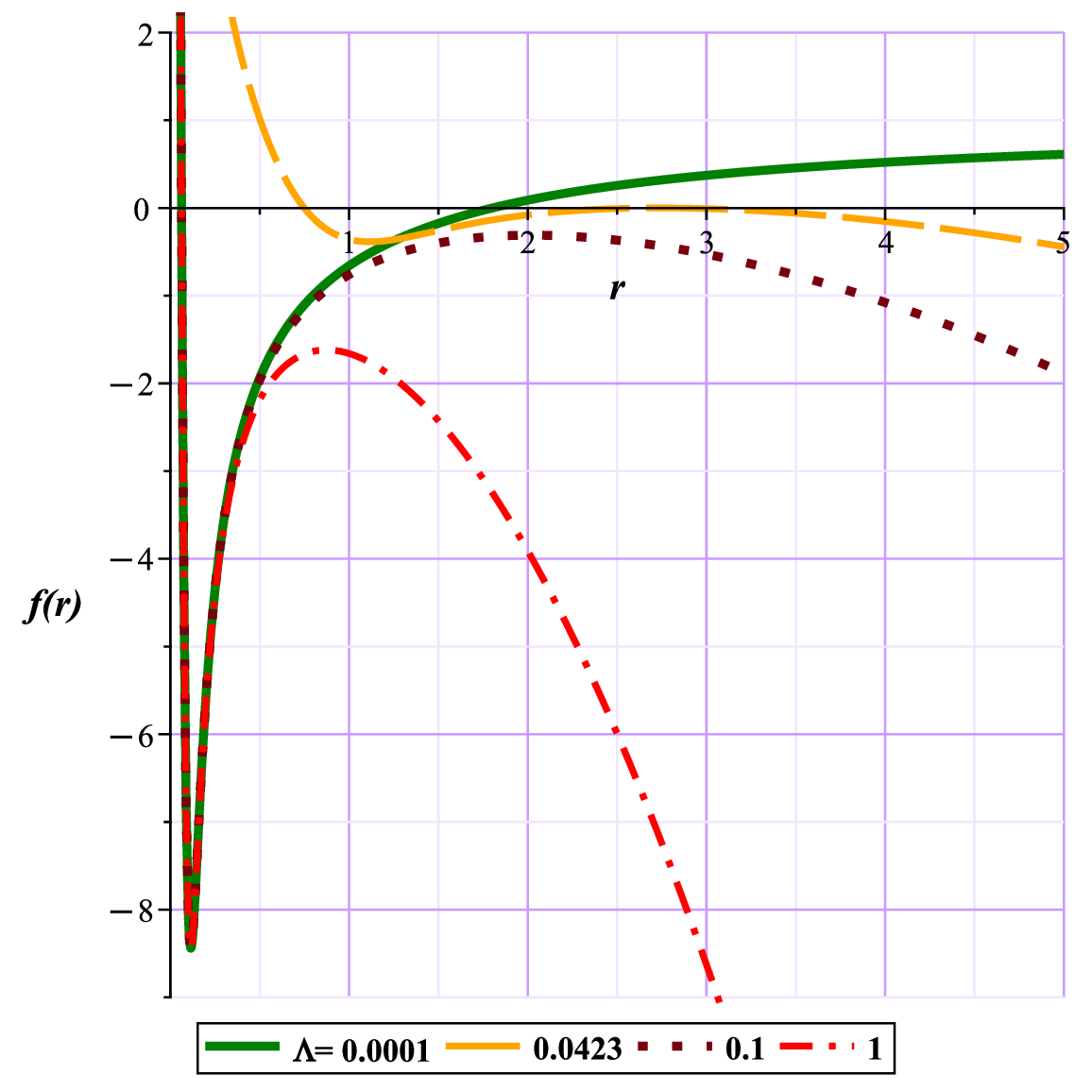}
 \label{12a}}
 \subfigure[]{
 \includegraphics[height=5.5cm,width=8cm]{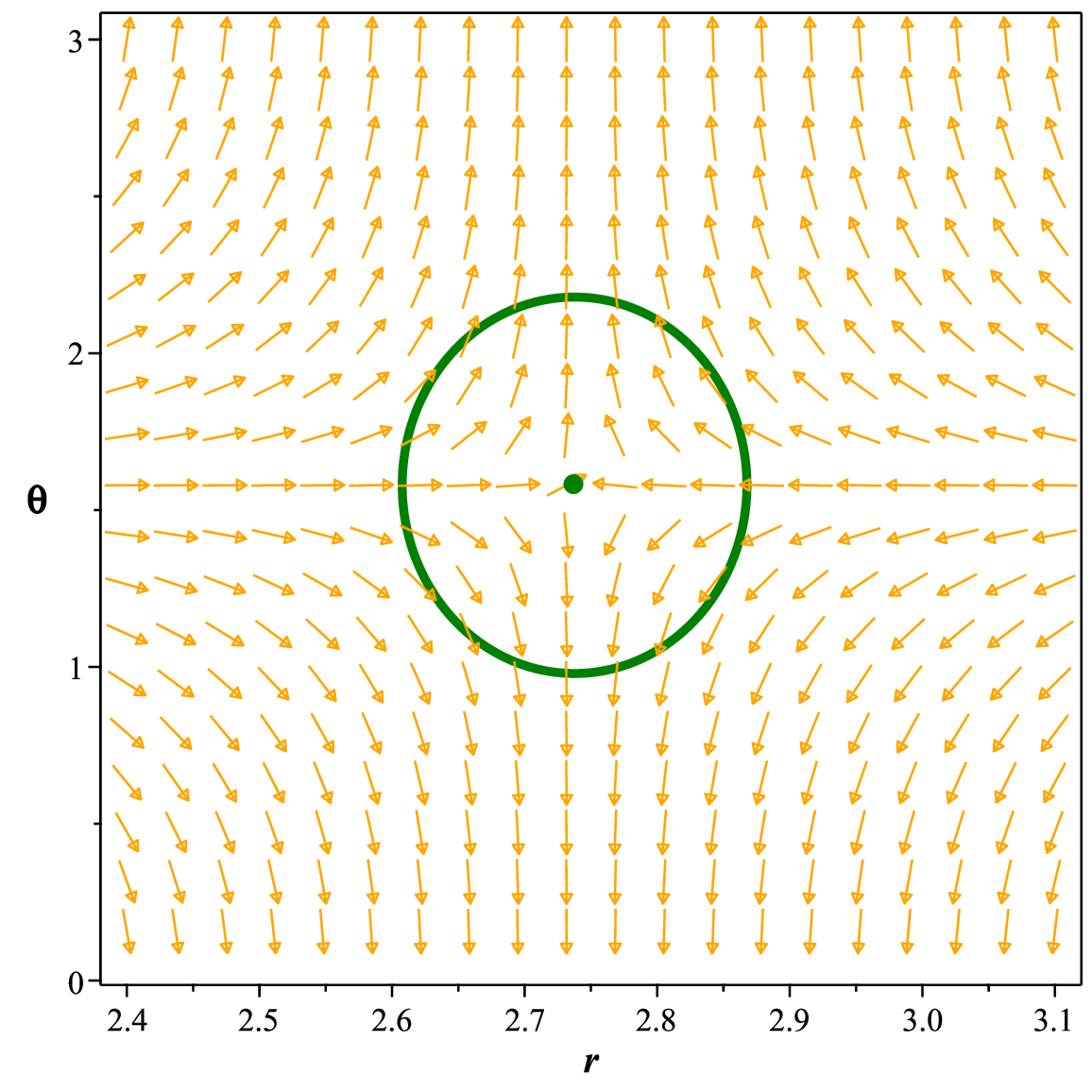}
 \label{12b}}
 \subfigure[]{
 \includegraphics[height=5.5cm,width=8cm]{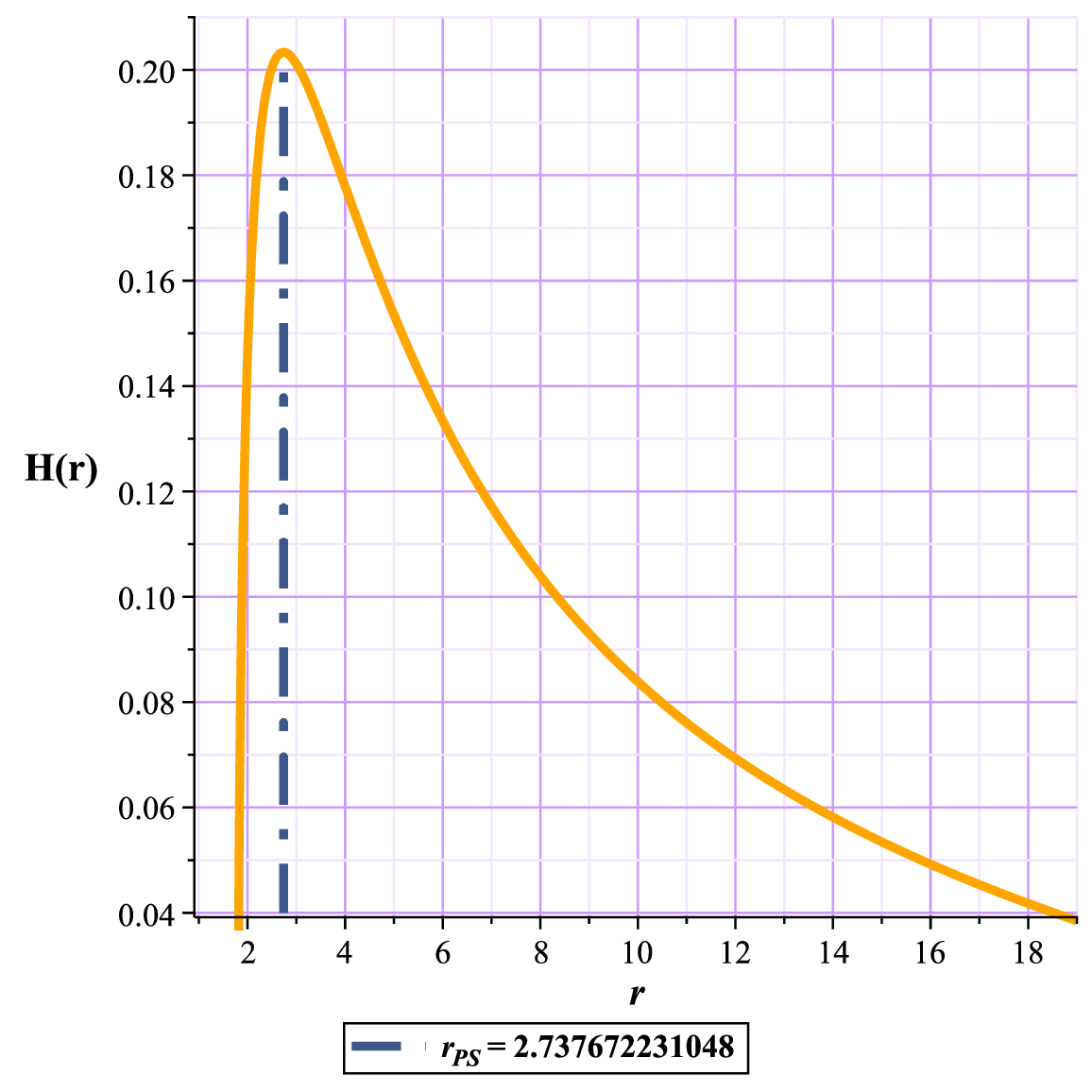}
 \label{12c}}
  \caption{\small{Fig.(12a)Metric function with different $\Lambda$ for NCBIBH Fig.(12b) The normal vector field $n$ in the $(r-\theta)$ plane. The photon spheres are located at  $(r,\theta)=(0.7031751797,1.57)$, $(r,\theta)=(2.737672231048,1.57)$ with respect to $(\Lambda=0.001, \Xi=0.001,b=0.5,q = 0.6, m = 1 )$, Figs.(12c) the topological potential H(r) for NCBIBH model }}
 \label{12}
\end{center}
 \end{figure}
\begin{center}
\begin{table}[H]
  \centering
\begin{tabular}{|p{3cm}|p{4cm}|p{4cm}|p{4cm}|}
  \hline
  \centering{NC Born-Infeld($\Lambda>0)$}  & \centering{Fix parametes} &\centering{Conditions}&*TTC\\[3mm]
   \hline
  \centering{unstable photon sphere} & $ \Xi=0.001,b=0.5,q = 0.6, m = 1$ & \centering{$0< \lambda \leq 0.0423$} &$-1$\\[3mm]
   \hline
  \centering{*Unauthorized area} & $ \Xi=0.001,b=0.5,q = 0.6, m = 1$ & \centering{$0.0423 < \Lambda $} &$nothing$\\[3mm]
   \hline
      \end{tabular}
   \caption{*Unauthorized region: is the region where the roots of $\phi$ equations become negative or imaginary in this region.\\ TTC: *Total Topological Charge\\}\label{1}
\end{table}
 \end{center}
As can be seen from table (3), dS models seem to have no naked singularity.
\section{Toward evidence for WGC }
The WGC is part of a broader research program known as the "Swampland Program,".
In fact, the WGC is an Effective Field Theory (a framework for describing a system using quantum field theory) detector, which seeks to distinguish between those which can be consistently embedded in a theory of quantum gravity (the "landscape": a chain of islands with valid EFTs) and those that cannot (the "swampland": a much larger ocean of swampland with inconsistent EFTs). The WGC provides criteria to identify theories that belong to the landscape \cite{42,43,44,45}.\\
This conjecture asserts that in any given effective field theory that has a gauge symmetry (like U (1)), there should exist a particle whose charge-to-mass ratio exceeds a certain threshold, which typically expressed as $ q/m > 1$. This ensures that gravity is the weakest force at sufficiently high energy scales. To understand this better, consider the following definitions:
The charged black holes can be sorted into three types:\\
1. Subextremal $(Q < M)$: The most common type.\\ 
2. Extremal $(Q = M)$: Similar to subextremal black holes, but here the inner (Cauchy) horizon and the outer (event) horizon merge into a single horizon.\\
3. Superextremal $(Q > M$): The rarest type, where the event horizon disappears entirely, leaving a naked singularity.\\
The charged particles can be sorted into two types:\\
1.Subextremal Particles: These particles have a charge-to-mass ratio less than 1.\\
2.Superextremal Particles: These have a charge-to-mass ratio greater than 1, meaning they are more "charged" than their mass would typically allow \cite{43}.\\
The relationship between the WGC and black holes is primarily centered around the stability and decay processes of extremal black holes\cite{45.1,46,47,48,49,50,51,52,53,54,55,56,57,58,59,60}. 
This is crucial because if the WGC is satisfied, it implies that extremal black holes can decay by emitting these superextremal particles, thus preventing the formation of naked singularities and preserving the cosmic censorship hypothesis.
If the WGC is violated, an extremal black hole could only shed its charge by emitting a subextremal particle, which would lead to a superextremal black hole and potentially introduce a naked singularity into spacetime, violating cosmic censorship. But why? If an extremal black hole can only emit subextremal particles, the decay process could be constrained. For instance, let’s say the black hole emits a subextremal particle:\\
1. The mass (energy) of the emitted particle decreases the total mass of the black hole.\\
2. Since the charge of the black hole is not being reduced sufficiently compared to the mass, the black hole could end up in a regime where its remaining mass is less than its charge.\\
3. This subsequently allows a black hole that does not already have the conditions to transform into the superextremal form to transform into the superextreme state, where $q/m > 1$.\\
Therefore, the WGC serves as a safeguard against such violations by ensuring that the decay processes of black holes are consistent with the principles of quantum gravity.\\
Consequently, based on the aforementioned discussions, one of the factors required for a black hole to undergo its natural collapse process until complete evaporation is the presence of superextremal particles. Clearly, the likelihood of such particles existing in subextremal and extremal black holes is zero or significantly lower compared to superextremal black holes. However, superextremal black holes, which generally exhibit naked singularities.\\
The significance lies in the fact that if we can identify models that exhibit the general behaviors of a normal black hole in a superextremal form, this would reinforce the validity and necessity of the WGC. This is because, on one hand, given that a superextremal black hole has a charge-to-mass ratio greater than one, the necessity of superextremal particles becomes crucial (as posited by the WGC). On the other hand, if such a black hole can exist without creating observable naked singularities, it indicates a stable balance of forces that constitute black hole mechanics.\\
To analyze a black hole with general behavior, various conditions such as the presence of an event horizon, WCCC, stability, thermodynamic properties, and more can be examined. One of the distinctive features of a black hole in its general state, besides the event horizon, is the appearance of a photon sphere. Given its ability to classify spacetime, as mentioned in this and previous articles, the photon sphere can be utilized to investigate the conditions of a model in the superextremal form. This allows us to determine the range of charge within which the model can remain in a black hole state, exhibiting a negative TTC and an unstable photon sphere, while still possessing the event horizon.\\Accordingly, our calculations indicate that the Charged NC model does not exhibit the structure of a black hole in the super-extremal form and instead appears as a naked singularity in this regime. However, when the nonlinear BI field is coupled to the AdS action, as observed in our previous calculations, not only does the model's accuracy and alignment with reality improve, 
but it also retains the ability to remain in a black hole form, even in the super-extremal state. In Fig.(13), we examine the system's tolerance for overcharging based on assumed values and observe that the model can withstand an increase in charge up to ( q/m = 1.178 ) for a unit mass. 
\begin{figure}[H]
 \begin{center}
 \subfigure[]{
 \includegraphics[height=6.5cm,width=8cm]{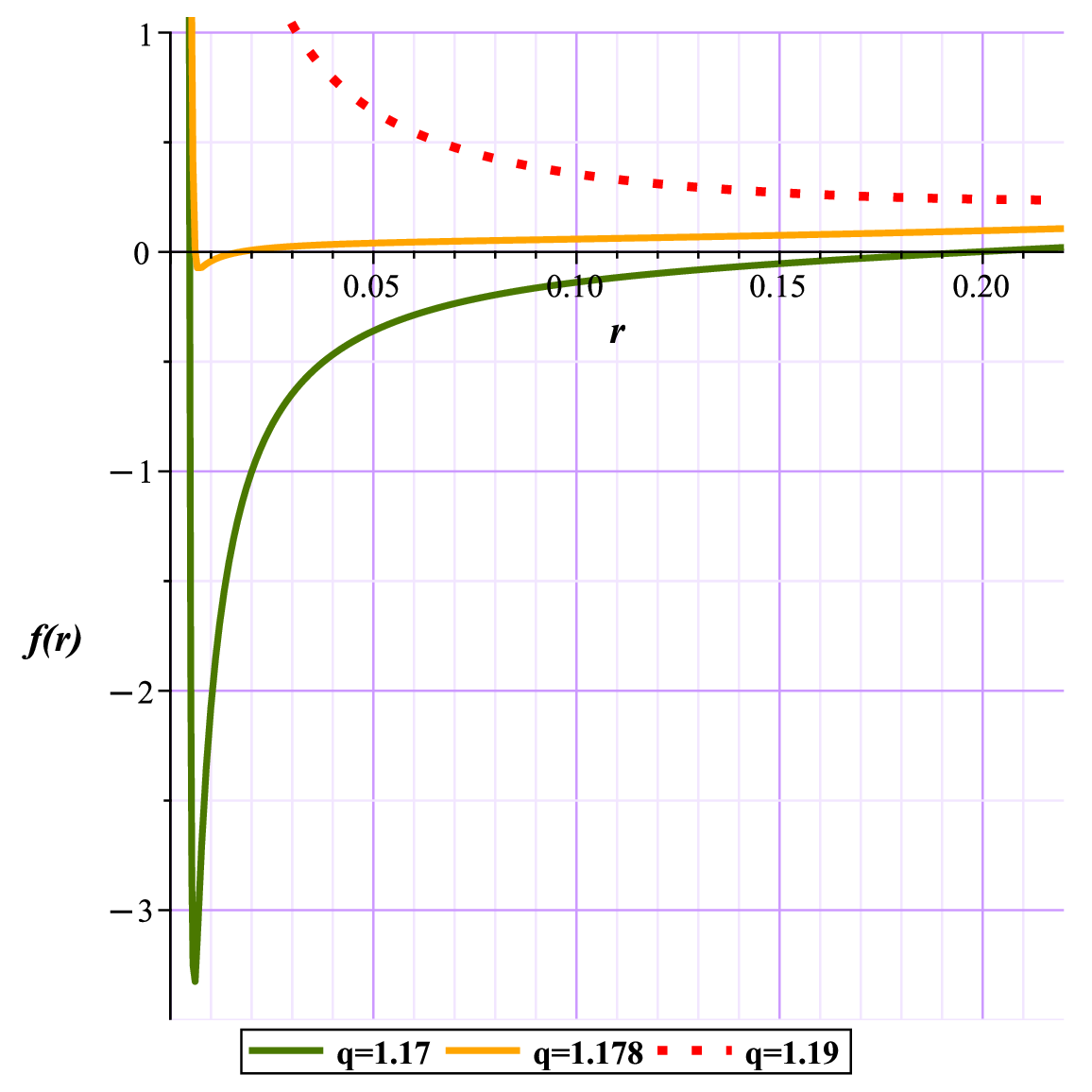}
 \label{1a}}
 \caption{\small{Metric function with m=1,b=0.4,$\Xi = 10^(-6),\Lambda=-1$,different $q$ for NCBIBH  }}
 \label{13}
\end{center}
\end{figure}
But is the existence of the event horizon alone enough to classify the model as a black hole? We address this question based on the photon sphere. As evident in Fig.(14), as long as we are away from the superextremality boundary (e.g., (q = 1.1)), the system follows its natural behavior, displaying black hole characteristics with (TTC = -1).
\begin{figure}[H]
 \begin{center}
 \subfigure[]{
 \includegraphics[height=5.5cm,width=8cm]{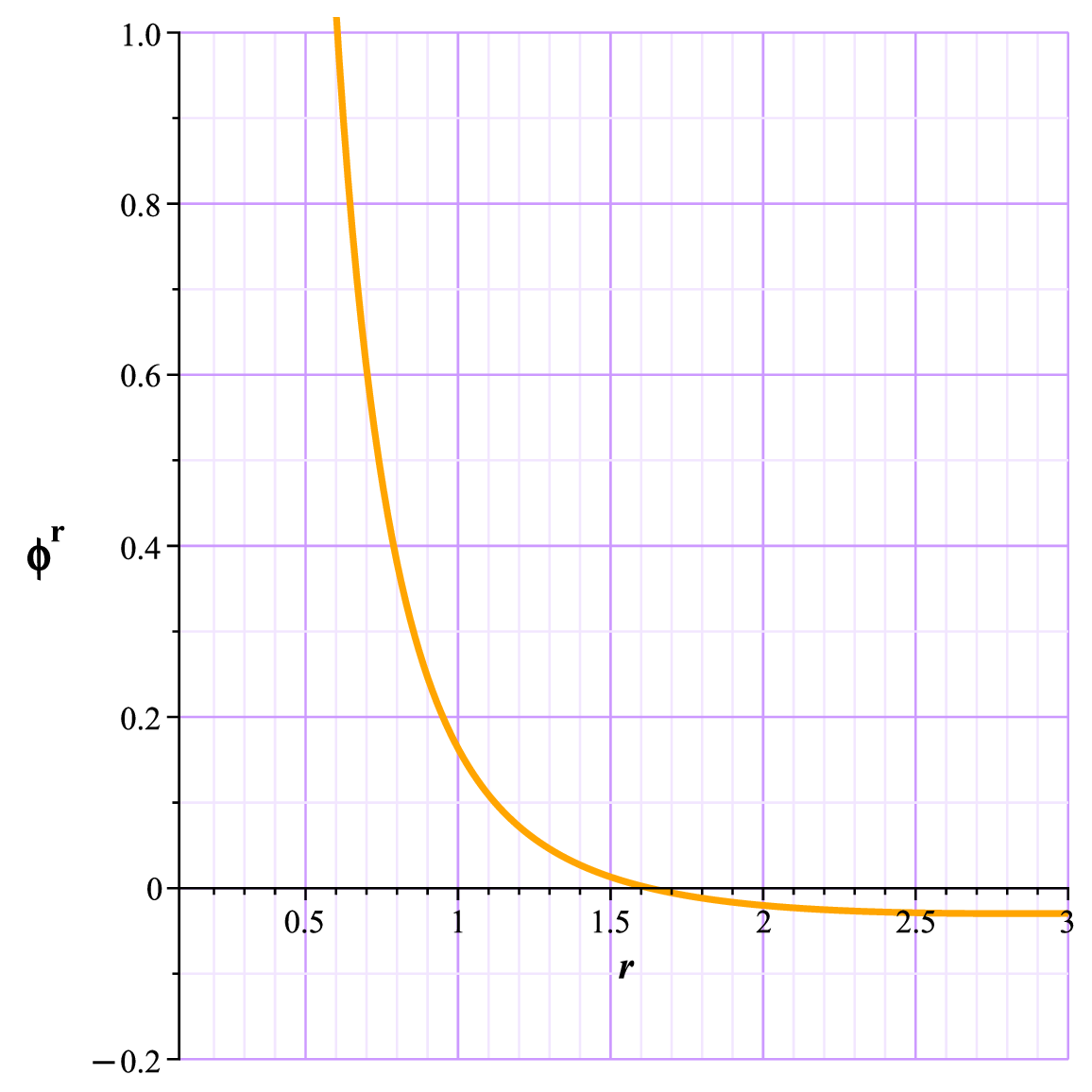}
 \label{10a}}
 \subfigure[]{
 \includegraphics[height=5.5cm,width=8cm]{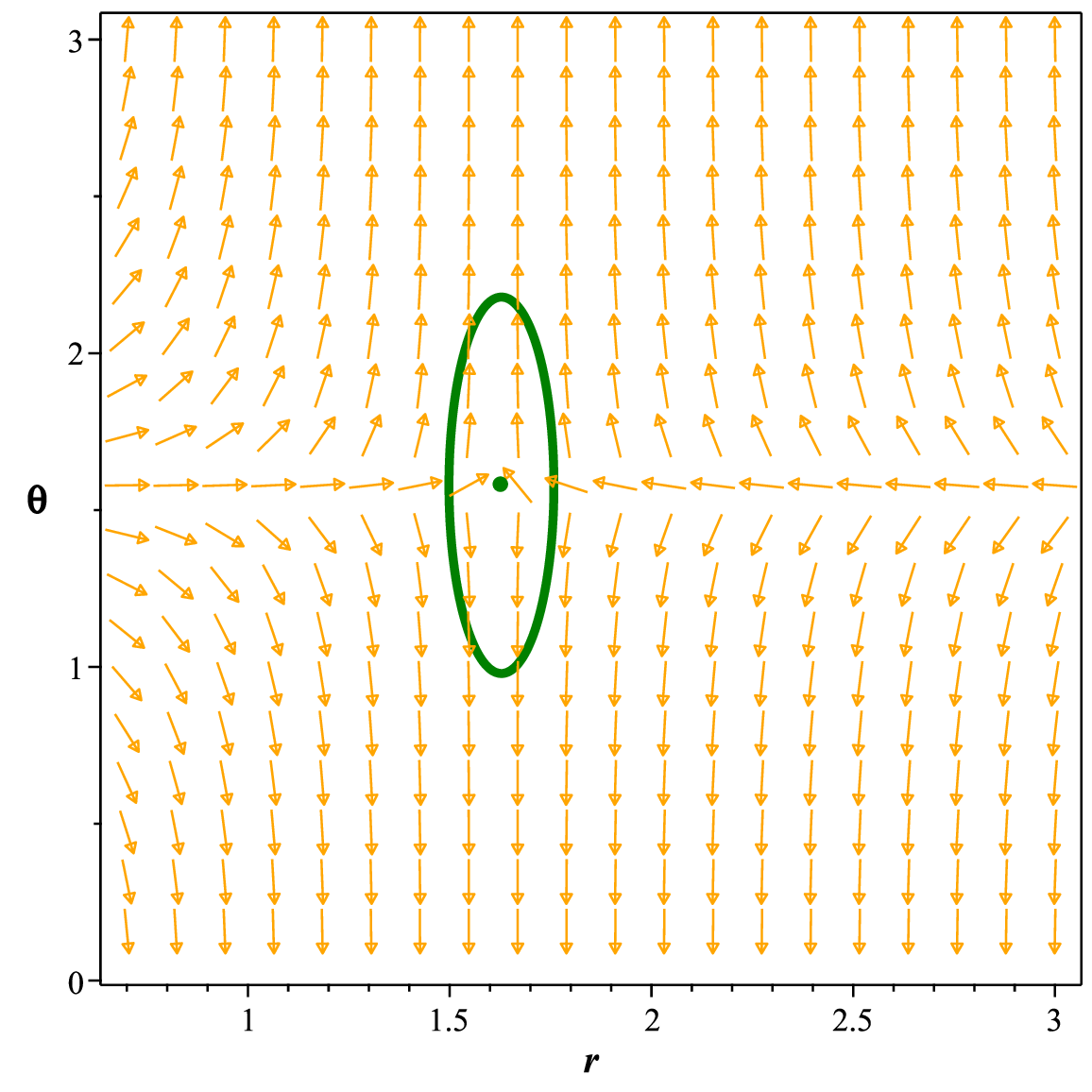}
 \label{10b}}
 
   \caption{\small{Fig (14a):Field component $\phi^{r}$  for NCBIBH with respect to $(\Lambda=-1,b=0.4, \Xi=10^-6,q = 1.1, m = 1 )$  , (14b): The photon sphere is located at $(r,\theta)=(1.62827889402,1.57)$  for NCBIBH  }}
 \label{2}
\end{center}
\end{figure}
However, as we approach the superextremality boundary, despite the model possessing an event horizon, the system does not exhibit  photon sphere. In this region, the normal vector, n, is both discontinuous and piecewise, and the $\phi^{r}$ lacks roots, as illustrated in Figures 15(a) and 15(b).
\begin{figure}[H]
 \begin{center}
 \subfigure[]{
 \includegraphics[height=5.5cm,width=8cm]{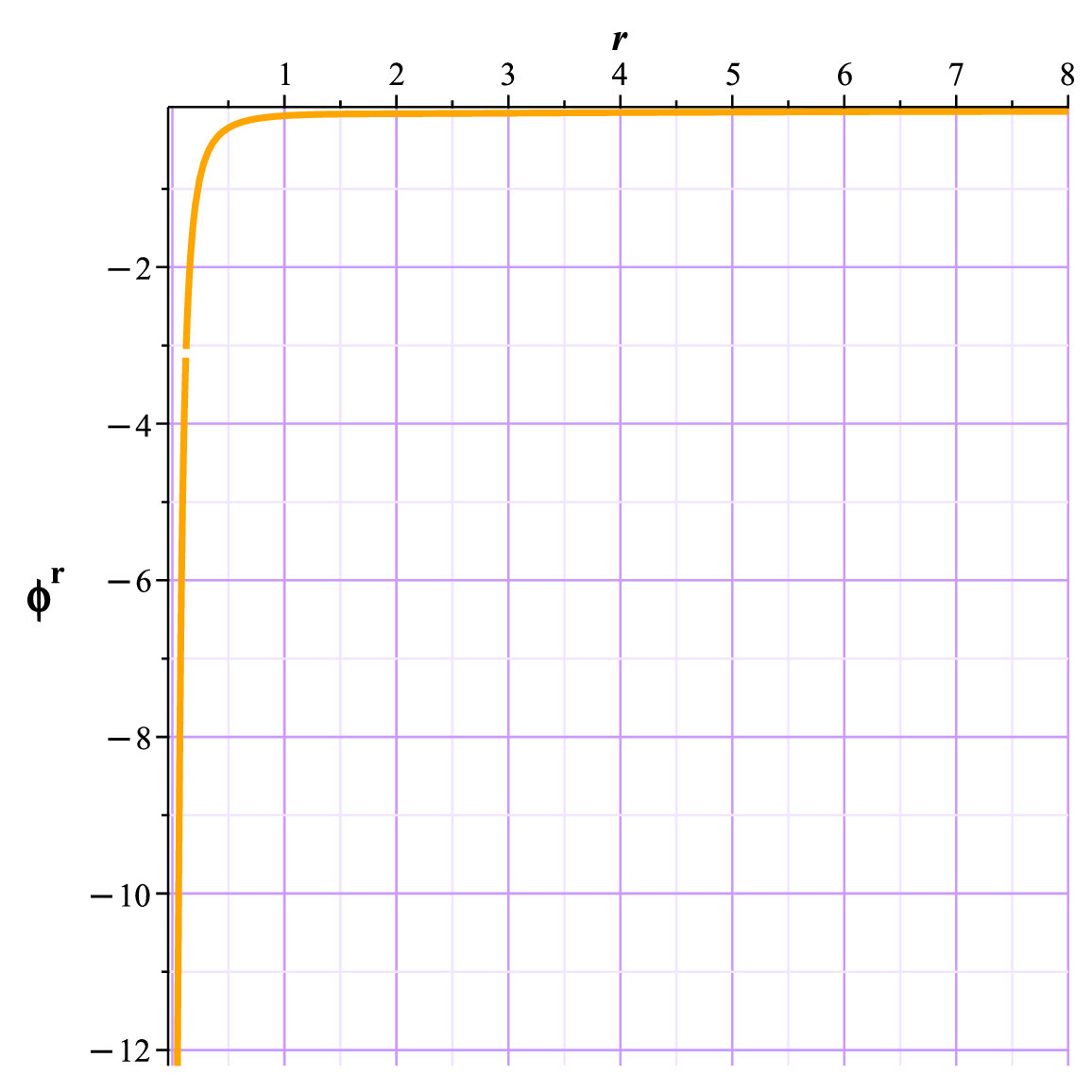}
 \label{10a}}
 \subfigure[]{
 \includegraphics[height=5.5cm,width=8cm]{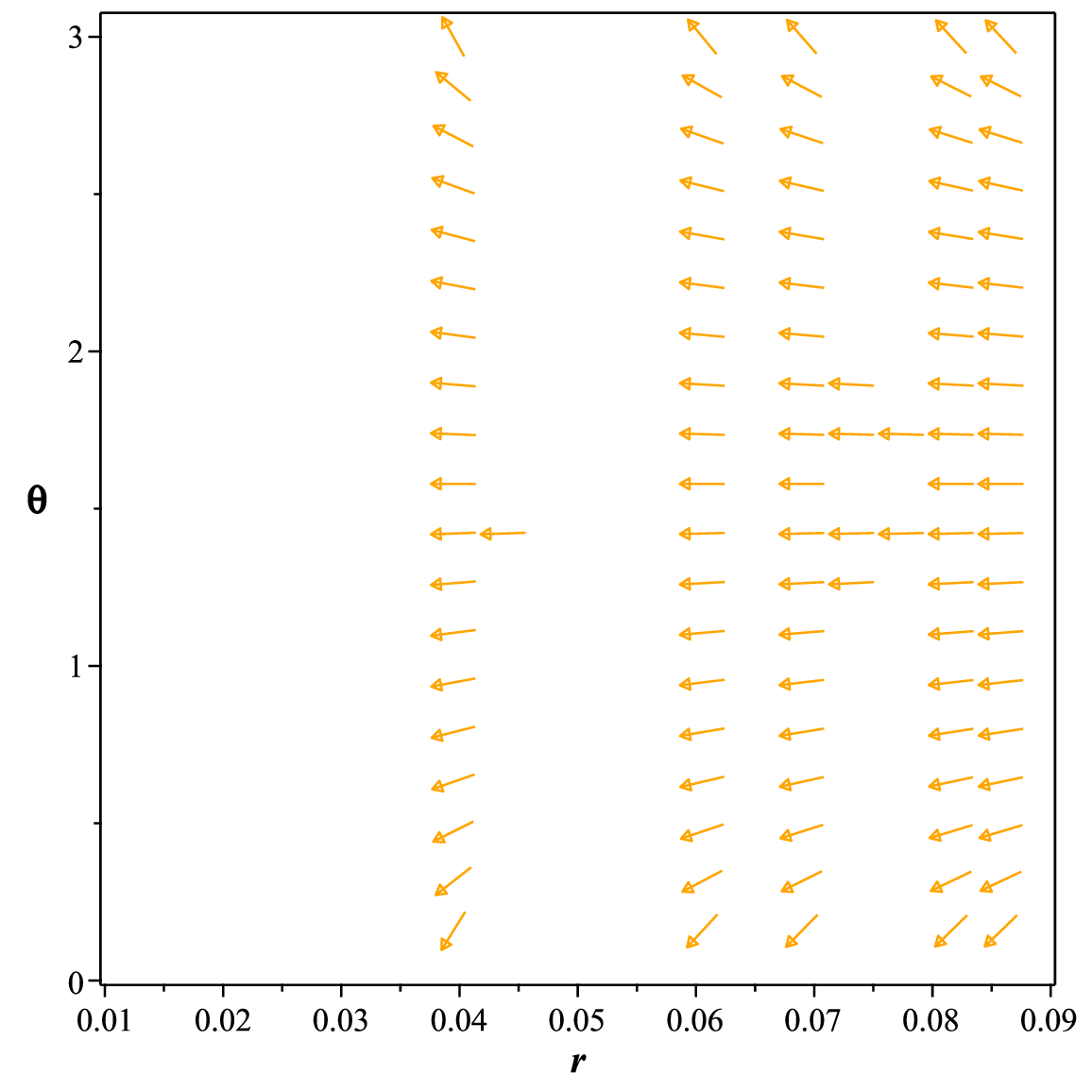}
 \label{10b}}
 
   \caption{\small{Fig (14a):Field component $\phi^{r}$  for NCBIBH with respect to $(\Lambda=-1,b=0.4, \Xi=10^-6,q = 1.178, m = 1 )$  , (14b): The normal vector field $n$ in the $(r-\theta)$ plane for NCBIBH  }}
 \label{2}
\end{center}
\end{figure}
Therefore, it seems that the model has lost one of its evidences of being a black hole.
\section{Conclusions}
In this series of articles, we explore NCBH. These black holes, characterized by a NC geometry where spacetime coordinates do not commute, aim to incorporate quantum effects into the fabric of spacetime to address the singularities and infinities that plague classical black hole solutions. The most significant difference arises from the modification of the spacetime metric due to non-commutativity, which replaces the Dirac delta function with a Gaussian distribution, effectively smearing the mass and eliminating point-like singularities.\\
With this choice, we pursued several objectives throughout the article. Firstly, we aimed to examine the impact of changing the mass distribution function from a point-like to an extended form on the geodesics of the black hole and demonstrate the significant role of the photon sphere as a powerful tool for studying black holes. Since our understanding of black holes, is primarily based on the study of waves and light rays emitted from them. Therefore, photon rings and photon spheres, as the edge of the shadow, play a crucial role in our comprehension of these phenomena.\\
In the Charged NC model, we clearly observe that altering the mass distribution function, despite eliminating the central singularity, does not significantly change our understanding of the black hole structure, particularly in comparison to the Reissner-Nordström model. However, the black hole structure imposes a constraint on the theta parameter, allowing us to observe a black hole structure only within the range $0 < \Xi < 0.2184$. Additionally, the study of the topological photon sphere indicates that the observable range of gravitational effects (black hole + naked singularity) for the $\Xi$ parameter in the model extends only up to (0.3273). The examination of TCOs also reveals the presence of both UTCOs and STCOs.
In the naked singularity form, we also examined the photon sphere and TCOs, noting the presence of two distinct regions of TCOs on either side of the stable and unstable photon spheres. An interesting point is the existence of a region of UTCOs behind the stable photon sphere and near the central singularity.\\Next, we explored a model where the nonlinear BI field is coupled to the NC action. We examined this model in two forms: with a negative and a positive cosmological constant. In the negative curvature case, the addition of the BI field to the model clearly enhances the accuracy and alignment with physics. On one hand, we know that the smaller the $\Xi$ parameter, the more realistic the results. On the other hand, our study shows that in this new state, the structure remains in the black hole form only up to $\Xi = 0.0611$, a significant reduction compared to the previous (0.2184). Interestingly, the range of gravitational influence remains constant at $0 < \Xi < 0.3273$, but the distribution indicates a more precise model. We see how the photon sphere effectively demonstrates the impact of parameters on each others.
The examination of TCOs in this model, both in the black hole and naked singularity forms, presents a standard framework similar to the previous model, although there are some differences in the results.\\
In the dS model, the results are quite remarkable. Our studies show that the model appears to be more influenced by the cosmological constant than by the BI parameter. For a conventional cosmological constant $\Lambda = 1$, changing the $\Xi $ or BI parameters does not play a significant role in the model, and the cosmological horizon does not appear. A crucial point to note is that the study of the effective potential and the photon sphere clearly indicates the instability of the model in the mentioned states. In other words, the study of the effective potential in this state can serve as an important test for predicting the model's status.
Our study showed that there is a critical limit for the cosmological constant in the dS state. Additionally, the model in the dS form still lacks a naked singularity format.\\
Finally, we turned our attention to the Weak Gravity Conjecture (WGC). In our previous two articles \cite{33,35}., we suggested that the photon sphere could be used as a tool to examine the conditions for the WGC in black hole models, which used in articles \cite{61,62}. In this article, we meticulously explained how the photon sphere in super-extremal models can serve as a powerful tool in understanding this conjecture. Additionally, with the help of the photon sphere, we demonstrated that, unlike the first model which could not exhibit the super-extremal state, the BI model not only possesses the super-extremal form but also retains its black hole structure. Furthermore, we showed how the photon sphere can assist in determining whether a super-extremal model, despite having an event horizon, qualifies as a black hole.

\end{document}